\documentclass[epj]{svjour}
\usepackage{graphics}

\def\half{\mbox{\small $\frac{1}{2}$}}

\def\nth{\mbox{\small $\frac{1}{n}$}}
\def\bvec#1{\ifmmode
\mathchoice{\mbox{\boldmath$\displaystyle\bf#1$}}
{\mbox{\boldmath$\textstyle\bf#1$}}
{\mbox{\boldmath$\scriptstyle\bf#1$}}
{\mbox{\boldmath$\scriptscriptstyle\bf#1$}}\else
{\mbox{\boldmath$\bf#1$}}\fi}

\begin{document}
\title{Statistical Models with Uncertain Error Parameters}
\author{Glen Cowan 
}                     
%
%
\institute{Physics Department, Royal Holloway, University of London, 
Egham, TW20 0EX, U.K.}
\date{Received: date / Revised version: date}
%
\abstract{
  In a statistical analysis in Particle Physics, nuisance parameters
  can be introduced to take into account various types of systematic
  uncertainties.  The best estimate of such a parameter is often
  modeled as a Gaussian distributed variable with a given standard
  deviation (the corresponding ``systematic error'').  Although the
  assigned systematic errors are usually treated as constants, in
  general they are themselves uncertain.
  A type of model is presented where the uncertainty in the assigned
  systematic errors is taken into account.  Estimates of the
  systematic variances are modeled as gamma distributed random
  variables.  The resulting confidence intervals show interesting and
  useful properties.  For example, when averaging measurements to
  estimate their mean, the size of the confidence interval increases
  for decreasing goodness-of-fit, and averages have reduced
  sensitivity to outliers.  The basic properties of the model are
  presented and several examples relevant for Particle Physics are
  explored.
%
%
\PACS{
      {02.50.Tt}{Inference methods}   \and
      {02.70.Rr}{General statistical methods}
     } 
} 
\maketitle

\section{Introduction}
\label{sec:intro}

Data analysis in Particle Physics is based on observation of a set of
numbers that can be represented by a (vector) random variable, here
denoted as $\bvec{y}$.  The probability of $\bvec{y}$ (or probability
density for continuous variables) can in general be written $P(\bvec{y}
| \bvec{\mu}, \bvec{\theta})$, where $\bvec{\mu}$ represents parameters
of interest and $\bvec{\theta}$ are nuisance parameters needed for the
correctness of the model but not of interest to the analyst.

The goal of the analysis is to carry out inference related to the
parameters of interest.  A procedure for doing this in the framework
of frequentist statistics using the profile likelihood function is
described in Sec.~\ref{sec:proflike}.  This involves using control
measurements with given standard deviations to provide information on
the nuisance parameters.
Here we will take the term ``systematic error'' to mean the standard
deviation of a control measurement itself.  The word ``error'' is used
in the sense defined here and not to mean, e.g., the unknown
difference between an inferred and true value.  The systematic errors
defined in this way should also not be confused with corresponding
systematic uncertainty in the estimate of the parameter of interest.

Often the values assigned to the systematic errors are themselves
uncertain.  This can be incorporated into the model by treating their
values as adjustable parameters and their estimates as random
variables.  A model is proposed in which the estimates of systematic
variances are treated as following a gamma distribution, whose mean
and width are set by the analyst to reflect the desired nominal value
and its relative uncertainty.

The confidence intervals that result from this type of model are found
to have interesting and useful properties.  For example, when
averaging measurements to estimate their mean, the size of the
confidence interval increases with decreasing goodness-of-fit, and
averages have reduced sensitivity to outliers.  The basic properties
of the model are presented and several types of examples relevant for
Particle Physics are explored.

The approach followed here is that of frequentist statistics, as this
is widely used in Particle Physics.  Models with elements similar to
the one proposed have been discussed in the statistics literature,
e.g., Refs.~\cite{bib:browne2006,bib:lange1989}.  Analogous Bayesian
procedures have been been investigated in Particle Physics
\cite{bib:dose2014,bib:dagostini1999,bib:cowan2006} and found to
produce results with qualitatively similar properties.

After reviewing parameter inference using the profile likelihood with
known systematic errors in Sec.~\ref{sec:proflike}, the model with
adjustable error parameters is presented in Sec.~\ref{sec:gammamod}
and its use in determining confidence intervals is discussed in
Sec.~\ref{sec:confint}.  In this paper two areas where such a model
can be applied are explored: a single Gaussian distribution
measurement in Sec.~\ref{sec:single} and the method of least squares
in Sec.~\ref{sec:ls}.  The issue of correlated systematic
uncertainties is discussed in Sec.~\ref{sec:corr} and conclusions are
given in Sec.~\ref{sec:conc}.

\section{Parameter inference using the profile likelihood and
the case of known systematic errors}
\label{sec:proflike}

Inference about a model's parameters is based on the likelihood
function $L(\bvec{\mu}, \bvec{\theta}) = P(\bvec{y} | \bvec{\mu},
\bvec{\theta})$.  More specifically one can construct a frequentist
test of values of the parameters of interest $\bvec{\mu}$ by using the
profile likelihood ratio (see, e.g., Ref.~\cite{asimov}),

\begin{equation}
\label{eq:proflike}
\lambda(\bvec{\mu}) = \frac{L(\bvec{\mu}, \hat{\hat{\bvec{\theta}}})}
{L(\hat{\bvec{\mu}}, \hat{\bvec{\theta}})} \,.
\end{equation}

\noindent Here in the denominator, $\hat{\bvec{\mu}}$ and
$\hat{\bvec{\theta}}$ represent the maximum-likelihood (ML) estimators
of $\bvec{\mu}$ and $\bvec{\theta}$, and $\hat{\hat{\bvec{\theta}}}$ are
the profiled values of $\bvec{\theta}$, i.e., the values of
$\bvec{\theta}$ that maximize the likelihood for a given value of
$\bvec{\mu}$.

Often the nuisance parameters are introduced to account for a
systematic uncertainty in the model.  
Their presence parameterizes the systematic uncertainty such that for
some point in the enlarged parameter space the model should be closer
to the truth.  Because of correlations between the estimators of the
parameters, however, the nuisance parameters result in a decrease in
sensitivity to the parameters of interest.  To counteract this
unwanted effect, one often includes into the set of observed
quantities additional measurements that provide information on the
nuisance parameters.

A simple and often used form of such control measurements involves
treating the best available estimates of the nuisance parameters
$\bvec{\theta} = (\theta_1, \ldots, \theta_N)$ as independent Gaussian
distributed values $\bvec{u} = (u_1, \ldots, u_N)$ with standard
deviations $\bvec{\sigma}_{\bvec{u}} = (\sigma_{u_1}, \ldots,
\sigma_{u_N})$.  In this way the full likelihood becomes

\begin{eqnarray}
\label{eq:pxu}
L(\bvec{\mu}, \bvec{\theta}) & = & 
P(\bvec{y}, \bvec{u} | \bvec{\mu}, \bvec{\theta}) = 
P(\bvec{y} | \bvec{\mu}, \bvec{\theta}) P(\bvec{u} | \bvec{\theta}) 
\nonumber \\*[0.3 cm]
& = & 
P(\bvec{y} | \bvec{\mu}, \bvec{\theta}) \, \prod_{i=1}^N 
\frac{1}{\sqrt{2 \pi} \sigma_{u_i}} e^{-(u_i - \theta_i)^2 / 2 \sigma_{u_i}^2 }
\,,
\end{eqnarray}

\noindent or equivalently the log-likelihood is 

\begin{equation}
\label{eq:lnpxu}
\ln L(\bvec{\mu}, \bvec{\theta}) = 
\ln P(\bvec{y} | \bvec{\mu}, \bvec{\theta}) 
- \frac{1}{2}  \sum_{i=1}^N \frac{(u_i - \theta_i)^2}{\sigma_{u_i}^2 }
+ C \,,
\end{equation}

\noindent where $C$ represents terms that do not depend on the
adjustable parameters of the problem and therefore can be 
dropped; in the
following such constant terms will usually not be written explicitly.

The log-likelihood in Eq.~(\ref{eq:lnpxu}) represents one of the most
widely used methods for taking account of systematic uncertainties in
Particle Physics analyses.  First nuisance parameters are introduced
into the model to parameterize the systematic uncertainty, and then
these parameters are constrained by means of control measurements.
The quad\-rat\-ic constraint terms in Eqs.~(\ref{eq:lnpxu}) correspond
to the case where the estimate $u_i$ of the parameter $\theta_i$ is
modeled as a Gaussian distributed variable of known standard deviation
$\sigma_{u_i}$.

In some problems one may have parameters $\eta_i$ that are
intrinsically positive with estimates $t_i$ modeled as following a
log-normal distribution.  The Gaussian model covers this case as well
by defining $\theta_i = \ln \eta_i$ and $u_i = \ln t_i$, so that $u_i$
is the corresponding Gaussian distributed estimator for $\theta_i$.

Often the estimates $u_i$ are the outcome of real control
measurements, and so the standard deviations $\sigma_{u_i}$ are
related to the corresponding sample size.  The control measurement
itself could, however, involve a number of uncertainties or arbitrary
model choices, and as a result the values of the $\sigma_{u_i}$ may
themselves be uncertain.

Gaussian modelling of the $u_i$ can be used even if the measurement
exists only in an idealized sense.  For example, the parameter
$\theta_i$ could represent a not-yet computed coefficient in a
perturbation series, and $u_i$ is one's best guess of its value (e.g.,
zero).  In this case one may try to estimate an appropriate
$\sigma_{u_i}$ by means of some recipe, e.g., by varying some aspects
of the approximation technique used to arrive at $u_i$.  For example,
in the case of prediction based on perturbation theory one may try
varying the renormalization scale in some reasonable range.  In such a
case the estimate of $\sigma_{u_i}$ results from fairly arbitrary
choices, and values that may differ by 50\% or even a factor of two
might not be unreasonable.

\section{Gamma model for estimated variances}
\label{sec:gammamod}

One can extend the model expressed by Eq.~(\ref{eq:pxu}) to account
for the uncertainty in the systematic errors by treating the
$\sigma_{u_i}$ as adjustable parameters.  The best estimates $s_i$ for
the $\sigma_{u_i}$ are regarded as measurements to be included in the
likelihood model.  The width of the distribution of the $s_i$ is set
by the analyst to reflect the appropriate uncertainty in the
$\sigma_{u_i}$.

The characterization of the ``error on the error'' is described in
Sec.~\ref{sec:erronerr}.  In Sec.~\ref{sec:proflikegam} the full
mathematical model is defined and the corresponding likelihood
profiled over the $\sigma_{u_i}$ is derived.  This is shown in
Sec.~\ref{sec:student} to be equivalent to a model in which the
estimates $u_i$ follow a Student's $t$ distribution.

\subsection{The relative error on the error}
\label{sec:erronerr}

In the model proposed here it is convenient to regard the variances
$\bvec{\sigma}^2_{\bvec{u}}$ as the parameters, and to take values $v_i
= s_i^2$ as their estimates.  There is a special case in which the
estimated variances $v_i$ will follow a chi-squared distribution,
namely, when $v_i$ is the sample variance of $n$ independent
observations of $u_i$, i.e.,

\begin{equation}
v_i = \frac{1}{n-1} \sum_{j=1}^n (u_{i,j} - \overline{u}_i)^2 \,,
\end{equation}

\noindent where $u_{i,j}$ is the $j$th observation of $u_i$ and
$\overline{u}_i = \nth \sum_{j=1}^n u_{i,j}$.  If the $u_{i,j}$ are
Gaussian distributed with standard deviations $\sigma_{u_i}$, then one
finds (see, e.g., Ref.~\cite{Kendall1}) that the statistic $(n-1) v_i
/ \sigma_{u_i}^2$ follows a chi-squared distribution for $n-1$ degrees
of freedom.  Furthermore, the chi-squared distribution for $n$ degrees
of freedom is a special case of the gamma distribution,

\begin{equation}
\label{eq:gammapdf}
f(v ; \alpha, \beta) = \frac{\beta^{\alpha}}{\Gamma(\alpha)} v^{\alpha - 1}
e^{- \beta v} \,, \quad \quad v \ge 0 \,,
\end{equation}

\noindent for parameter values $\alpha = n/2$ and $\beta = 1/2$.  The
mean and variance are related to the parameters $\alpha$ and $\beta$
by $E[v] = \alpha / \beta$ and $V[v] = \alpha/\beta^2$. Therefore if
$(n-1)v_i/\sigma_{u_i}^2$ follows a chi-square distribution with $n-1$
degrees of freedom, then $v_i$ follows a gamma distribution with

\begin{eqnarray}
\label{eq:alphan}
\alpha_i & = & \frac{n-1}{2} \,, \\*[0.3 cm] 
\label{eq:betan}
\beta_i & = & \frac{n-1}{2\sigma_{u_i}^2} \,.
\end{eqnarray}

In general the analyst will not base the estimate $v_i$ on $n$
observations of $u_i$ but rather on different types of information,
such as related control measurements or approximate theoretical
predictions.  The analyst must then set the width of the distribution
of $v_i$ to reflect the appropriate level of uncertainty in the
estimate of $\sigma_{u_i}^2$.

For $v_i = s_i^2$, using error propagation gives to first order

\begin{equation}
\label{eq:sigmav}
\frac{\sigma_{v_i}}{E[v_i]} \approx 2 \frac{\sigma_{s_i}}{E[s_i]} \,.
\end{equation}

\noindent To characterize the width of the gamma distribution
we define

\begin{equation}
\label{eq:rsigmav}
r_i \equiv \frac{1}{2} \frac{\sigma_{v_i}}{E[v_i]} = 
\frac{1}{2} \frac{\sigma_{v_i}}{\sigma_{u_i}^2} \,.
\end{equation}

\noindent From Eq.~(\ref{eq:sigmav}) one sees that to first
approximation $r_i \approx \sigma_{s_i} / E[s_i]$ and thus we can
think of these factors as representing the relative uncertainty in the
estimate of the systematic error.  The parameters $r_i$ will be
referred to as the ``error on the error''.  A more accurate relation
between $r_i$ as defined here and the quantity
$\sigma_{s_i} / E[s_i]$ is given in Appendix~\ref{sec:rdef}.

Using the expectation value of the gamma distribution $E[v_i] =
\alpha_i / \beta_i$ and its variance $V[v_i] = \alpha_i / \beta_i^2$,
we can relate the values $r_i$ supplied by the analyst and the
$\sigma_{u_i}$ to $\alpha_i$ and $\beta_i$ by

\vspace{-0.2 cm}

\begin{eqnarray}
\label{eq:alphai}
\alpha_i & = & \frac{1}{4r_i^2}  \,, \\*[0.3 cm]
\label{eq:betai}
\beta_i & = & \frac{1}{4 r_i^2 \sigma_{u_i}^2}  \,.
\end{eqnarray}

Figure~\ref{fig:gamma}(a) shows the gamma distribution for $\sigma_u =
1$ and several values of $r$ and \ref{fig:gamma}(b) shows the
corresponding distribution of $s = \sqrt{v}$.  More details on the
distribution of $s$ and its properties are given in
App.~\ref{sec:rdef}.  The assumption of a gamma distribution is not
unique but represents nevertheless a reasonable and flexible
expression of uncertainty in the $\sigma_{u_i}$.  Moreover it will be
shown that by using the gamma distribution one finds a very simple
procedure for incorporating uncertain systematic errors into the
model.

\setlength{\unitlength}{1.0 cm}
\renewcommand{\baselinestretch}{0.9}
\begin{figure*}[htbp]
\begin{picture}(10.0,6.5)
\put(1,-0.2)
{\includegraphics{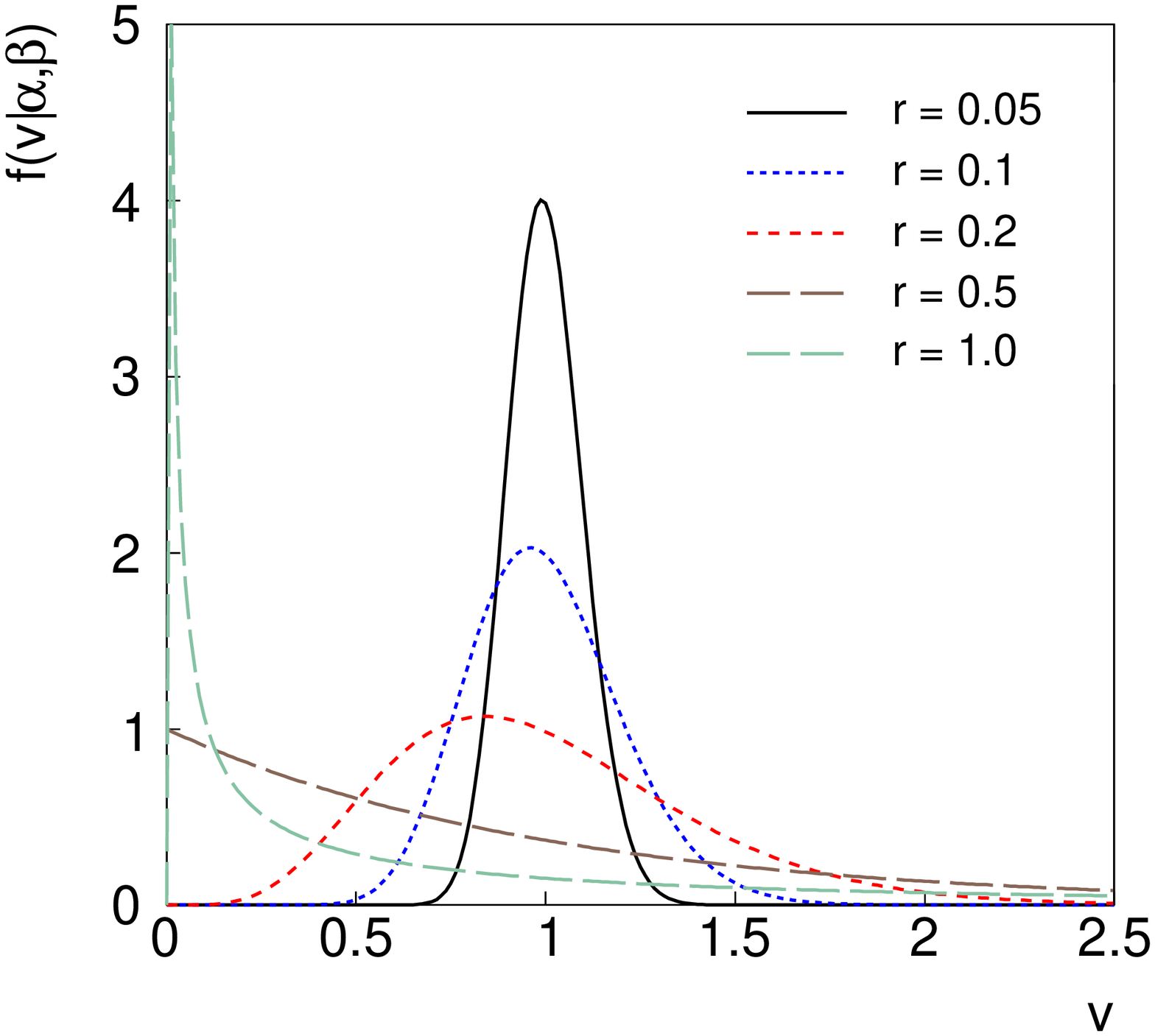}}
\put(10,-0.2)
{\includegraphics{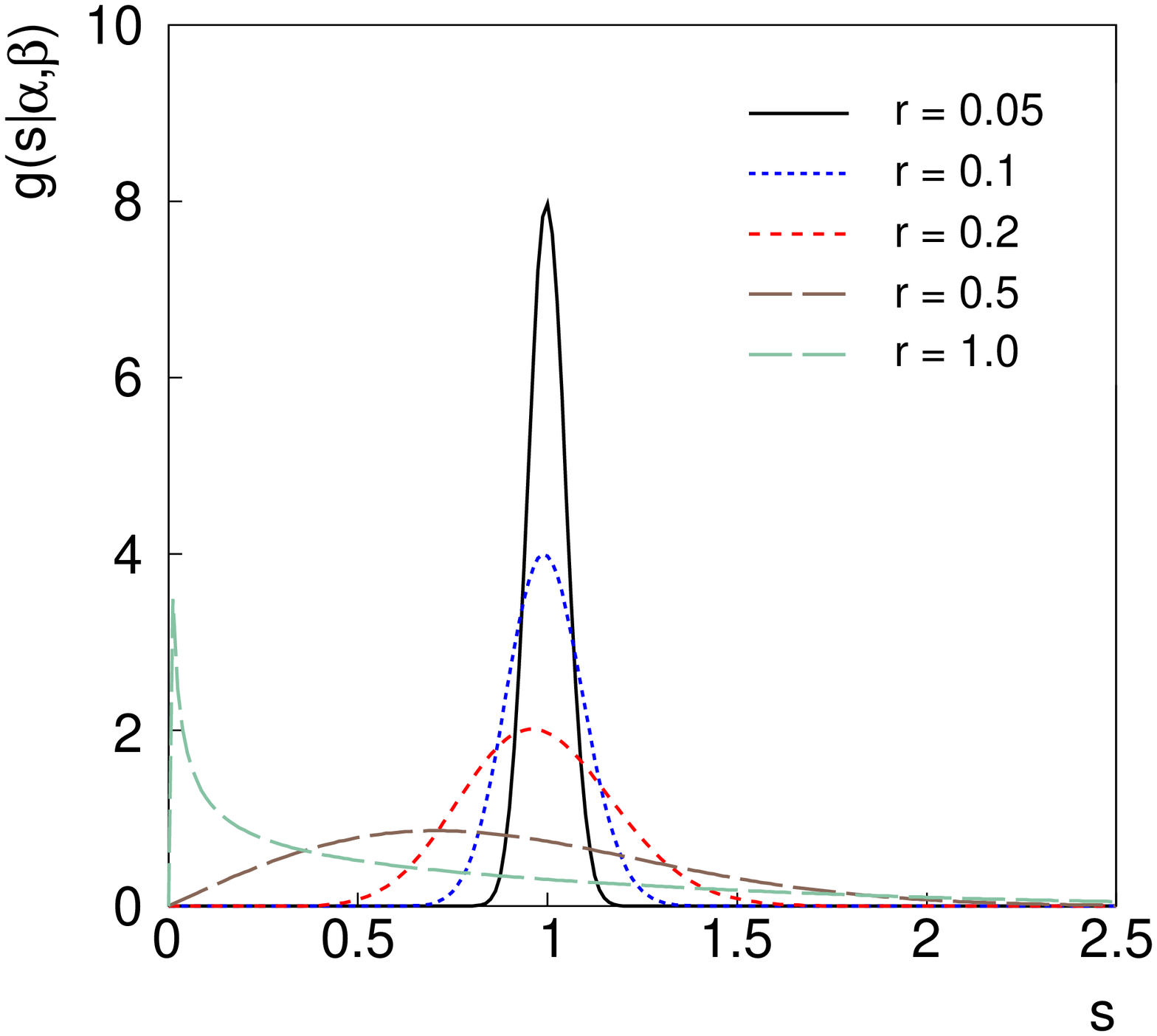}}
\put(8.,5.5){(a)}
\put(17,5.5){(b)}
\end{picture}
\caption{\small Plots of  (a) the gamma distribution
of the estimated variance $v$ and (b)
the Nakagami distribution for the estimated standard deviation
$s = \sqrt{v}$ for several values of the parameter $r$ (see text).} 
\label{fig:gamma}
\end{figure*}
\renewcommand{\baselinestretch}{1}
\small\normalsize

Using Eq.~(\ref{eq:alphan}) to connect the relative uncertainty $r_i$
to the effective number of measurements $n$ gives $n = 1 + 1/2r_i^2$.
A relevant special case is $n=2$, sometimes called the problem of
``two-point systematics'', where one has two estimates $u_{i,1}$ and
$u_{i,2}$ of a parameter $\theta_i$.  This gives

\vspace{-0.2 cm}

\begin{eqnarray}
\label{eq:ubari2pt}
\hat{\theta}_i &  = & 
\overline{u}_i  =  \half (u_{i,1} + u_{i,2}) \,, \\*[0.3 cm]
\label{eq:si2pt}
s_i & = & \frac{| u_{i,1} - u_{i,2} |}{\sqrt{2}} \,, \\*[0.3 cm]
\label{eq:ri2pt}
r_i & = & 1 / \sqrt{2} \,.
\end{eqnarray}


It will be assumed in this paper that the analyst is able to assign
meaningful values for the error-on-the-error parameters $r_i$.
The procedure for doing this will involve elicitation of expert
knowledge from those who assigned the systematic errors and will in
general vary depending on the experiment.  One may want to regard a
subset of the measurements as having a certain common $r$ which could
be fitted from the data, but we do not investigate this possibility
further here.


The proposed model thus makes two important assumptions.  First, the
control measurements are taken to be independent and Gaussian
distributed.  As mentioned in Sec.~\ref{sec:proflike}, the Gaussian
$u_i$ can be extended to an alternative distribution if it can be
related to a Gaussian by a transformation.  Second, the estimates of
the variances of the $u_i$ are gamma distributed.  Both assumptions
are reasonable but neither is a perfect description in practice, and
thus the resulting inference could be subject to corresponding
systematic uncertainties.  Nevertheless the proposed model will in
general be an improvement over the widely used Gaussian assumption for
$u_i$ with fixed variances.  In addition, the choice of the gamma
distribution leads to important simplifications in mathematical
expressions needed for inference, as shown in
Sec.~\ref{sec:proflikegam} below.

\subsection{ Likelihood for the gamma model}
\label{sec:proflikegam}

By treating the estimated variances $\bvec{v} = (v_1, \ldots, v_N)$ as
independent gamma distributed random variables, the full likelihood
function becomes

\begin{eqnarray}
\label{eq:fullL}
& & \! \! \! \! \! \! \! \! \! \!\! \! \! \! \! \! \! 
L(\bvec{\mu}, \bvec{\theta}, \bvec{\sigma}_{\bvec{u}}^2)  =
P(\bvec{y} | \bvec{\mu}, \bvec{\theta}) \nonumber \\*[0.3 cm]
& &  \! \! \! \! \! \! \! \! \! \! 
\times \prod_{i=1}^N   \frac{1}{\sqrt{2 \pi} \sigma_{u_i}} 
e^{-(u_i - \theta_i)^2 / 2 \sigma_{u_i}^2 } \,
\frac{\beta_{i}^{\alpha_i}}{\Gamma(\alpha_i)}
v_i^{\alpha_i - 1} e^{-\beta_i v_i} \,.
\end{eqnarray}

\noindent By using Eqs.~(\ref{eq:alphai}) and (\ref{eq:betai}) to
relate the parameters $\alpha_i$ and $\beta_i$ to $\sigma_{u_i}^2$ and
$r_i$ one finds, up to additive terms that are independent of the
parameters, the log-likelihood

\begin{eqnarray}
\label{eq:lnLfull}
& & \! \! \! \! \! \! 
\ln L(\bvec{\mu}, \bvec{\theta}, \bvec{\sigma}_{\bvec{u}}^2) = 
\ln P(\bvec{y} | \bvec{\mu}, \bvec{\theta}) 
\nonumber \\*[0.3 cm]
& & \! \! \! \!\! \! \! \!\! \!
- \frac{1}{2}
\sum_{i=1}^N \left[ 
\frac{(u_i - \theta_i)^2}{\sigma_{u_i}^2} 
+ \left( 1 + \frac{1}{2 r_i^2} \right) \ln \sigma_{u_i}^2 
+ \frac{v_i}{2 r_i^2 \sigma_{u_i}^2} \right] \,.
\end{eqnarray}

By setting the derivatives of $\ln L$ with respect to the
$\sigma_{u_i}^2$ to zero for fixed $\bvec{\theta}$ and $\bvec{\mu}$
one finds the profiled values

\begin{equation}
\label{eq:sigmau2hathat}
\widehat{\widehat{\sigma^2}}_{u_i} = 
\frac{v_i + 2 r_i^2 (u_i - \theta_i)^2}{1 + 2 r_i^2} \,.
\end{equation}

\noindent Using these for the $\sigma_{u_i}^2$ gives the profile
likelihood with respect to the systematic variances, but which still
depends on $\bvec{\theta}$ as well as the parameters of interest
$\bvec{\mu}$.  After some manipulation it can be written up to
constant terms as

\begin{eqnarray}
\label{eq:proflikmub}
& & \! \! \! \! \! \! \! \! \! \! \! \! \! \! \! \! \! \!  
 \ln L^{\prime}(\bvec{\mu}, \bvec{\theta}) 
=  \ln L (\mu, \bvec{\theta}, \widehat{\widehat{\bvec{\sigma}^2}}_{\bvec{u}}) 
\nonumber \\*[0.3 cm]
& = & \ln P(\bvec{y} | \bvec{\mu}, \bvec{\theta}) 
\nonumber \\*[0.3 cm]
& - & \frac{1}{2} \sum_{i=1}^N 
\left( 1  + \frac{1}{2 r_i^2} \right) \ln \left[
1 + 2 r_i^2 \frac{(u_i - \theta_i)^2}{v_i} 
\right] \,.
\end{eqnarray}

\noindent Some intermediate steps in the derivation of
Eq.~(\ref{eq:proflikmub}) are given in App.~\ref{sec:proflnL}.  In the
limit where all of the $r_i$ are small, the estimates $v_i$ are very
close to their expectation values $\sigma_{u_i}^2$.  Making this
replacement and expanding the logarithmic terms to first order one
recovers the quadratic terms as in Eq.~(\ref{eq:lnpxu}).

\subsection{Derivation of profile likelihood from Student's $t$ distribution}
\label{sec:student}

An equivalent derivation of the profile likelihood
(\ref{eq:proflikmub}) can be obtained by first defining

\begin{equation}
\label{eq:zidef}
z_i \equiv \frac{u_i - \theta_i}{\sqrt{v_i}} \,.
\end{equation}

\noindent As $u_i$ follows a Gaussian with mean $\theta_i$ and
standard deviation $\sigma_{u_i}$, and $v_i$ follows a gamma
distribution with mean $\sigma_{u_i}^2$ and standard deviation
$\sigma_{v_i} = 2 r_i^2 \sigma_{u_i}^2$, one can show (see, e.g.,
Ref.~\cite{Kendall1}) that $z_i$ follows a Student's $t$ distribution,

\begin{equation}
\label{eq:student}
f(z_i | \nu_i) = \frac{\Gamma\left( \frac{\nu_i + 1}{2} \right)}
{\sqrt{\nu_i \pi} \Gamma(\nu_i/2)}
\left( 1 + \frac{z_i^2}{\nu_i} \right)^{- \frac{\nu_i + 1}{2}} \,,
\end{equation}

\noindent with a number of degrees of freedom 

\begin{equation}
\label{eq:studentdof}
\nu_i = \frac{1}{2 r_i^2} \,.
\end{equation}

\noindent By constructing the likelihood $L(\bvec{\mu}, \bvec{\theta})$
as the product of $P(\bvec{y} | \bvec{\mu}, \bvec{\theta})$ and Student's
$t$ distributions,

\begin{equation}
L(\bvec{\mu}, \bvec{\theta}) = P(\bvec{y} | \bvec{\mu}, \bvec{\theta}) \, 
\prod_{i=1}^N \frac{\Gamma\left( \frac{\nu_i + 1}{2} \right)}
{\sqrt{\nu_i \pi} \Gamma(\nu_i/2)}
\left( 1 + \frac{z_i^2}{\nu_i} \right)^{- \frac{\nu_i + 1}{2}} \,,
\end{equation}

\noindent one obtains the same log-likelihood as given by $\ln
L^{\prime}$ from Eq.~(\ref{eq:proflikmub}).  That is, the same model
results if one replaces the estimates $v_i$ by constants
$\sigma_{u_i}^2$, but still takes the $z_i$ to follow a Student's $t$
distribution, with $u_i = \theta_i + \sigma_{u_i} z_i$.  Thus in the
following we can drop the prime in the profile log-likelihood
(\ref{eq:proflikmub}) and regard this equivalently as the
log-likelihood resulting from a model where the control measurements
are distributed according to a Student's $t$.  In the limit where $r_i
\rightarrow 0$ and thus the number of degrees of freedom $\nu_i
\rightarrow \infty$, the Student's $t$ distribution becomes a Gaussian
(see, e.g., Ref.~\cite{Kendall1}), and the corresponding term in the
log-likelihood becomes quadratic in $u_i - \theta_i$, as in
Eq.~(\ref{eq:lnpxu}).

\section{Estimators and confidence regions from profile likelihood}
\label{sec:confint}

\sloppy The ML estimators are found by maximizing the full
$\ln~L(\bvec{\mu}, \bvec{\theta}, \bvec{\sigma}^2_{\bvec{u}})$ with
respect to all of the parameters, which is equivalent to maximizing
the profile likelihood with respect to $\bvec{\mu}$ and
$\bvec{\theta}$.  In this way the statistical uncertainties due to
both the estimated biases $u_i$ as well as their estimated variances
$v_i$ are incorporated into the variances of the estimators for the
parameters of interest $\hat{\bvec{\mu}}$.

Consider for example the case of a single continuous parameter of
interest $\mu$.  Having found the estimator $\hat{\mu}$, one could
quantify its statistical precision by using the standard deviation
$\sigma_{\hat{\mu}}$.  The covariance matrix for all of the estimated
parameters can to first approximation be found from the inverse of the
matrix of second derivatives of $\ln L$ (see, e.g.,
Refs.~\cite{bib:Cowan98,bib:PDG}).  From this we extract the variance
of the estimator of the parameter of interest $\mu$, i.e.,
$V[\hat{\mu}] = \sigma^2_{\hat{\mu}}$.  The presence of the nuisance
parameters in the model will in general inflate $\sigma_{\hat{\mu}}$,
which reflects the corresponding systematic uncertainties.

But $\sigma_{\hat{\mu}}$ is by construction a property of the model
and not of a particular data set.  One may want, however, to report a
measure of uncertainty along with the estimate $\hat{\mu}$ that
reflects the extent to which the data values are consistent with the
hypothesized model, and therefore $\sigma_{\hat{\mu}}$ is not suitable
for this purpose.  We will show below, however, that a confidence
region can be constructed that has this desired property.

In general to find a confidence region (or for a single parameter a
confidence interval) one tests all values of $\bvec{\mu}$ with a test
of size $\alpha$ for some fixed probability $\alpha$.  Those values of
$\bvec{\mu}$ that are not rejected by the test constitute a confidence
region with confidence level $1 - \alpha$.  To determine the critical
region of the test of a given $\bvec{\mu}$ one can use a test statistic
based on the profile likelihood ratio

\begin{equation}
\label{eq:tmudef}
t_{\bvec{\mu}} = - 2 \ln \lambda(\bvec{\mu}) 
= -2 \ln \frac{L(\bvec{\mu}, \hat{\hat{\bvec{\theta}}})}
{L(\hat{\bvec{\mu}}, \hat{\bvec{\theta}})} \,.
\end{equation}

\noindent The critical region of a test of $\bvec{\mu}$ corresponds to
the region of data space having probability content $\alpha$ with
maximal $t_{\bvec{\mu}}$.  Equivalently, provided $t_{\bvec{\mu}}$ can
be treated as continuous, the $p$-value of a hypothesized point in
parameter space $\bvec{\mu}$ is

\begin{equation}
\label{eq:pvaldef}
p_{\bvec{\mu}} = \int_{t_{\bvec{\mu},{\rm obs}}}^{\infty} 
f(t_{\bvec{\mu}} | \bvec{\mu}, \bvec{\theta}, \bvec{\sigma}^2_{\bvec{u}}) \,
dt_{\bvec{\mu}} = 1 - F(t_{\bvec{\mu},{\rm obs}}) \,,
\end{equation}

\noindent where $t_{\bvec{\mu},{\rm obs}}$ is the observed value of
$t_{\bvec{\mu}}$ and $F$ is the cumulative distribution of
$t_{\bvec{\mu}}$.  That is, we define the region of data space even
less compatible with the hypothesis than what was observed to
correspond to $t_{\bvec{\mu}} > t_{\bvec{\mu},{\rm obs}}$.

The boundary of the confidence region corresponds to the values of
$\bvec{\mu}$ where $p_{\bvec{\mu}} = \alpha$.  Solving
Eq.~(\ref{eq:pvaldef}) for the test statistic gives

\begin{equation}
\label{eq:tmufpmu}
t_{\bvec{\mu}} = F^{-1}(1 - p_{\bvec{\mu}}) \,,
\end{equation}

\noindent where here $t_{\bvec{\mu}}$ refers to the value observed, and
$F^{-1}$ is the quantile of $t_{\bvec{\mu}}$.  The statistic
$t_{\bvec{\mu}}$ is also defined in terms of the likelihood through
Eqs.~(\ref{eq:proflike}) and (\ref{eq:tmudef}), and by using
$p_{\bvec{\mu}} = \alpha$ one finds that the boundary of the confidence
region is given by

\begin{equation}
\label{eq:confintbound}
\ln L(\bvec{\mu}, \hat{\hat{\bvec{\theta}}}) = 
\ln L(\hat{\bvec{\mu}}, \hat{\bvec{\theta}}) 
- \frac{1}{2} F^{-1}(1 - \alpha) \,.
\end{equation}

To find the $p$-values and thus determine the boundary of the
confidence region one needs the distribution $f(t_{\bvec{\mu}} |
\bvec{\mu}, \bvec{\theta}, \bvec{\sigma}_{\bvec{u}}^2)$.  According to
Wilks' theorem \cite{bib:Wilks}, for $M$ parameters of interest
$\bvec{\mu} = (\mu_1, \ldots, \mu_M)$ the statistic $t_{\bvec{\mu}}$
should follow a chi-squared distribution for $M$ degrees of freedom in
the asymptotic limit, which here corresponds to the case where the
distributions of all ML estimators are Gaussian.  To the extent that
this approximation holds we may identify the quantile $F^{-1}$ in
Eq.~(\ref{eq:confintbound}) with $F_{\chi^2_M}^{-1}$, the chi-squared
quantile for $M$ degrees of freedom.

If it is further assumed that the log-likelihood can be well
approximated by a quadratic function about its maximum, then one finds
asymptotically (see, e.g., Ref.~\cite{Kendall2}) that

\begin{equation}
\label{eq:confregfromlnl}
\ln L(\bvec{\mu}, \hat{\hat{\bvec{\theta}}}) =
\ln L(\hat{\bvec{\mu}}, \hat{\bvec{\theta}}) - 
\frac{1}{2} 
(\bvec{\mu} - \hat{\bvec{\mu}})^T V^{-1} (\bvec{\mu} - \hat{\bvec{\mu}}) \,,
\end{equation}

\noindent where $V_{ij} = \mbox{cov}[\hat{\mu}_i, \hat{\mu}_j]$ is the
covariance matrix for the parameters of interest.  This equation says
that the confidence region is a hyper-ellipsoid of fixed size centred
about $\hat{\bvec{\mu}}$.  For example, for a single parameter $\mu$
one finds that the endpoints

\begin{equation}
\label{eq:muplusminus}
\mu_{\pm} = \hat{\mu} \pm \sigma_{\hat{\mu}} \left[ 
F_{\chi^2_1}^{-1}(1 - \alpha) \right]^{1/2} 
\end{equation}

\noindent give the central confidence interval with confidence level
$1 - \alpha$.  For a probability content corresponding to plus or
minus one standard deviation about the centre of a Gaussian, i.e., $1
- \alpha$ = 68.3\%, one has $F_{\chi^2_1}^{-1}(1 - \alpha) = 1$, which
gives the well-known result that the interval of plus or minus one
standard deviation about the estimate is asymptotically a 68.3\% CL
central confidence interval.

The relations (\ref{eq:confregfromlnl}) and (\ref{eq:muplusminus})
depend, however, on a quadratic approximation of the log-likelihood.
In the model where the $\bvec{\sigma}_{\bvec{u}}$ are treated as
adjustable, the profile log-likelihood is given by
Eq.~(\ref{eq:proflikmub}), which contains terms that are logarithmic
in $(u_i - \theta_i)^2$, and not just the quadratic terms that appear
in Eq.~(\ref{eq:lnpxu}).  As a result the relation
(\ref{eq:confregfromlnl}) is only a good approximation in the limit of
small $r_i$, which is not always valid in the present problem.

We can nevertheless use Eq.~(\ref{eq:confintbound}) assuming a
chi-squared distribution for $t_{\bvec{\mu}}$ as a first approximation
for confidence regions.  We will see in the examples below that these
have interesting properties that already capture the most important
features of the model.  If higher accuracy is required then Monte
Carlo methods can be used to determine the distribution of
$t_{\bvec{\mu}}$.  Alternatively we can modify the statistic so that
its distribution is closer to the asymptotic form; this is explored
further in Sec.~\ref{sec:bartlett}.

\section{Single-measurement model}
\label{sec:single}

To investigate the asymptotic properties of the profile likelihood
ratio it is useful to examine a simple model with a single measured
value $y$ following a Gaussian with mean $\mu$ and standard deviation
$\sigma$.  The parameter of interest is $\mu$ and we treat the
variance $\sigma^2$ as a nuisance parameter, which is constrained by
an independent gamma-distributed estimate $v$.  Thus the likelihood is
given by

\begin{eqnarray}
\label{eq:toyL}
\! \! \! \! \! \! \! \! \! \! 
L(\mu, \sigma^2) & = & f(y, v | \mu, \sigma^2) 
\nonumber \\*[0.3 cm]
\! \! \! \! \! \! \! \! \! \! 
& = & 
\frac{1}{\sqrt{2 \pi \sigma^2}} e^{- (y - \mu)^2/2 \sigma^2} 
\, \frac{\beta^{\alpha}}{\Gamma(\alpha)} v^{\alpha - 1} e^{ - \beta v} \,.
\end{eqnarray}

\noindent As before we set the parameters $\alpha$ and $\beta$ of the
gamma distribution so that $E[v] = \sigma^2$ and so that from
Eq.~(\ref{eq:rsigmav}) the standard deviation of $v$ is $\sigma_v = 2
r \sigma^2$, where $r$ characterizes the relative error on the error.
This gives

\begin{eqnarray}
\label{eq:alpha}
\alpha & = & \frac{1}{4r^2} \,, \\*[0.3 cm]
\label{eq:beta}
\beta & = & \frac{1}{4 r^2 \sigma^2} \,.
\end{eqnarray}

\noindent The goal is to construct a confidence interval for $\mu$ by
using the profile likelihood ratio

\begin{equation}
\label{eq:toyPL}
\lambda(\mu) = \frac{L(\mu, \widehat{\widehat{\sigma^2}}(\mu))}
{L(\hat{\mu}, \widehat{\sigma^2})} \,.
\end{equation}

\noindent The log-likelihood is

\begin{equation}
\label{eq:toylnL}
\ln L(\mu, \sigma^2) = - \frac{1}{2} \frac{(y - \mu)^2}{\sigma^2}
- \left( \frac{1}{2} + \frac{1}{4 r^2} \right) \ln \sigma^2
- \frac{v}{4 r^2 \sigma^2} + C \,,
\end{equation}

\noindent where $C$ represents constants that do not depend on $\mu$
or $\sigma^2$.  From this we find the required estimators

\begin{eqnarray}
\hat{\mu} & = & y \,, \\*[0.3 cm]
\widehat{\sigma^2} & = & \frac{v}{1 + 2 r^2} \,, \\*[0.3 cm]
\widehat{\widehat{\sigma^2}}(\mu) & = & 
\frac{v + 2 r^2 (y - \mu)^2}{1 + 2 r^2} \,.
\end{eqnarray}

\noindent With these ingredients we find the following simple 
expression for the statistic $t_{\mu} = - 2 \ln \lambda(\mu)$,

\begin{eqnarray}
\label{eq:tmudef2}
t_{\mu} = \left( 1 + \frac{1}{2 r^2} \right)
\ln \left[ 1 + 2 r^2 \frac{(y - \mu)^2}{v} \right] \,.
\end{eqnarray}

According to Wilks' theorem \cite{bib:Wilks}, the distribution
$f(t_{\mu} | \mu)$ should, in the large-sample limit, be chi-squared
for one degree of freedom.  The large-sample limit corresponds to the
situation where estimators for the parameters become Gaussian, which
in this case means $r \ll 1$.
%
%

The behaviour of the distribution of $t_{\mu}$ for nonzero $r$ is
illustrated in Fig.~\ref{fig:tmudist}, which shows the distributions
from data generated according to a Gaussian of mean $\mu = 0$,
standard deviation $\sigma = 1$ and values of $r = 0.01$, $0.2$, $0.4$
and $0.6$.  The case of $r = 0.01$ approximates the situation where
the relative uncertainty on $\sigma$ is negligibly small.  One can see
that greater values of $r$ lead to an increasing departure of the
distribution from the asymptotic form.

\setlength{\unitlength}{1.0 cm}
\renewcommand{\baselinestretch}{0.9}
\begin{figure*}[htbp]
\begin{picture}(10.0,10.5)
\put(2,5.2) {\includegraphics{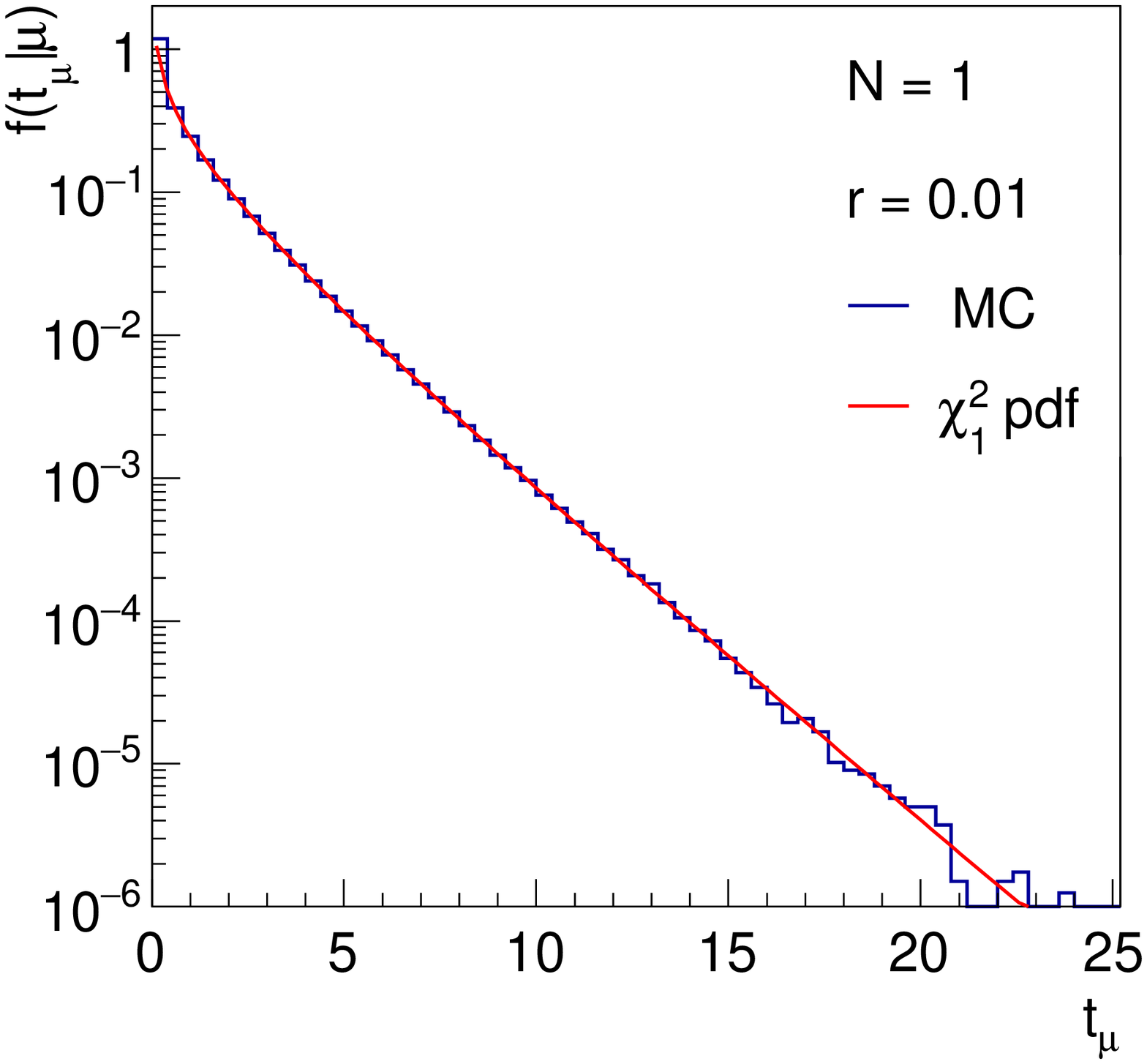}}
\put(10,5.2) {\includegraphics{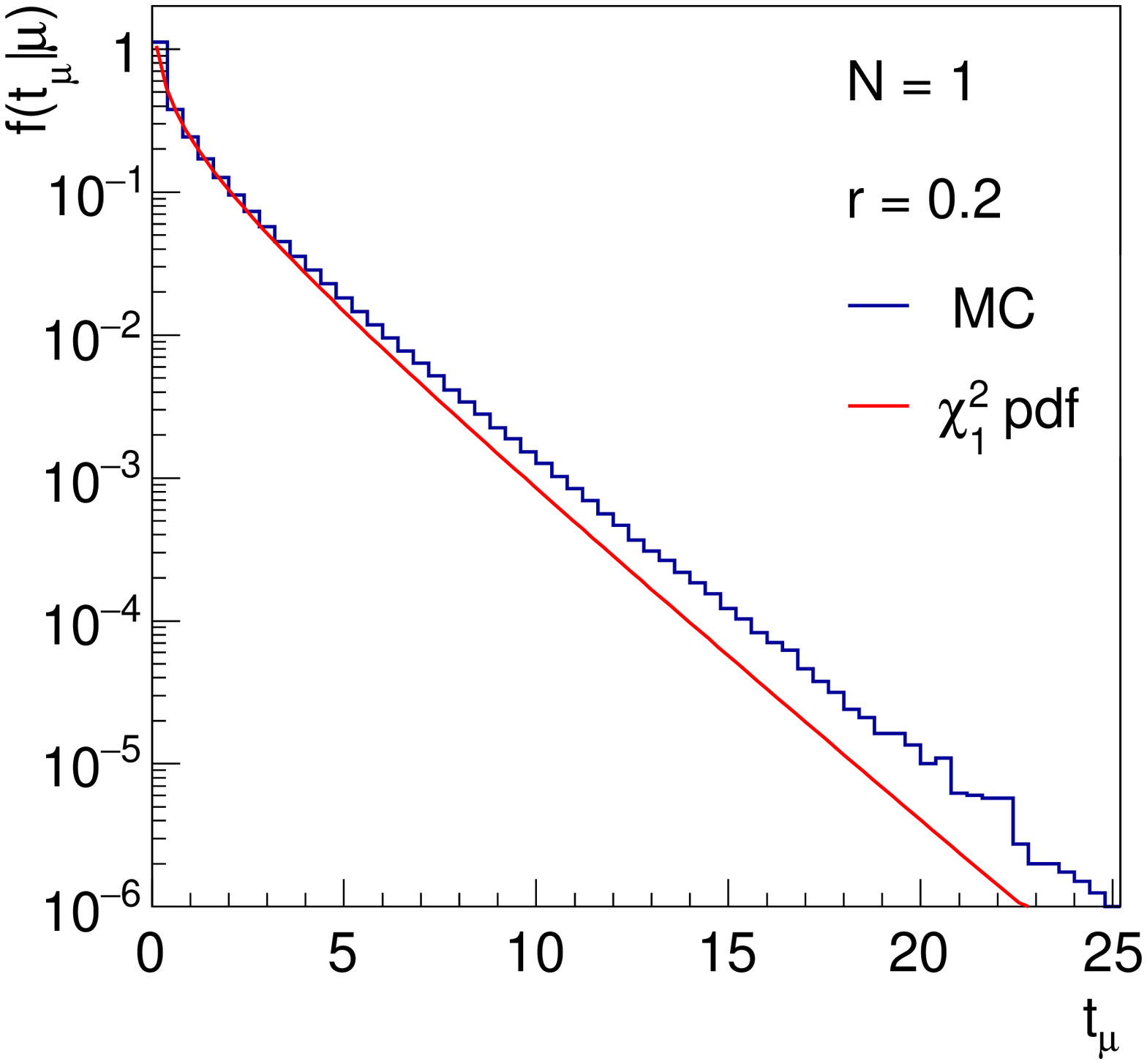}}
\put(2,0.0) {\includegraphics{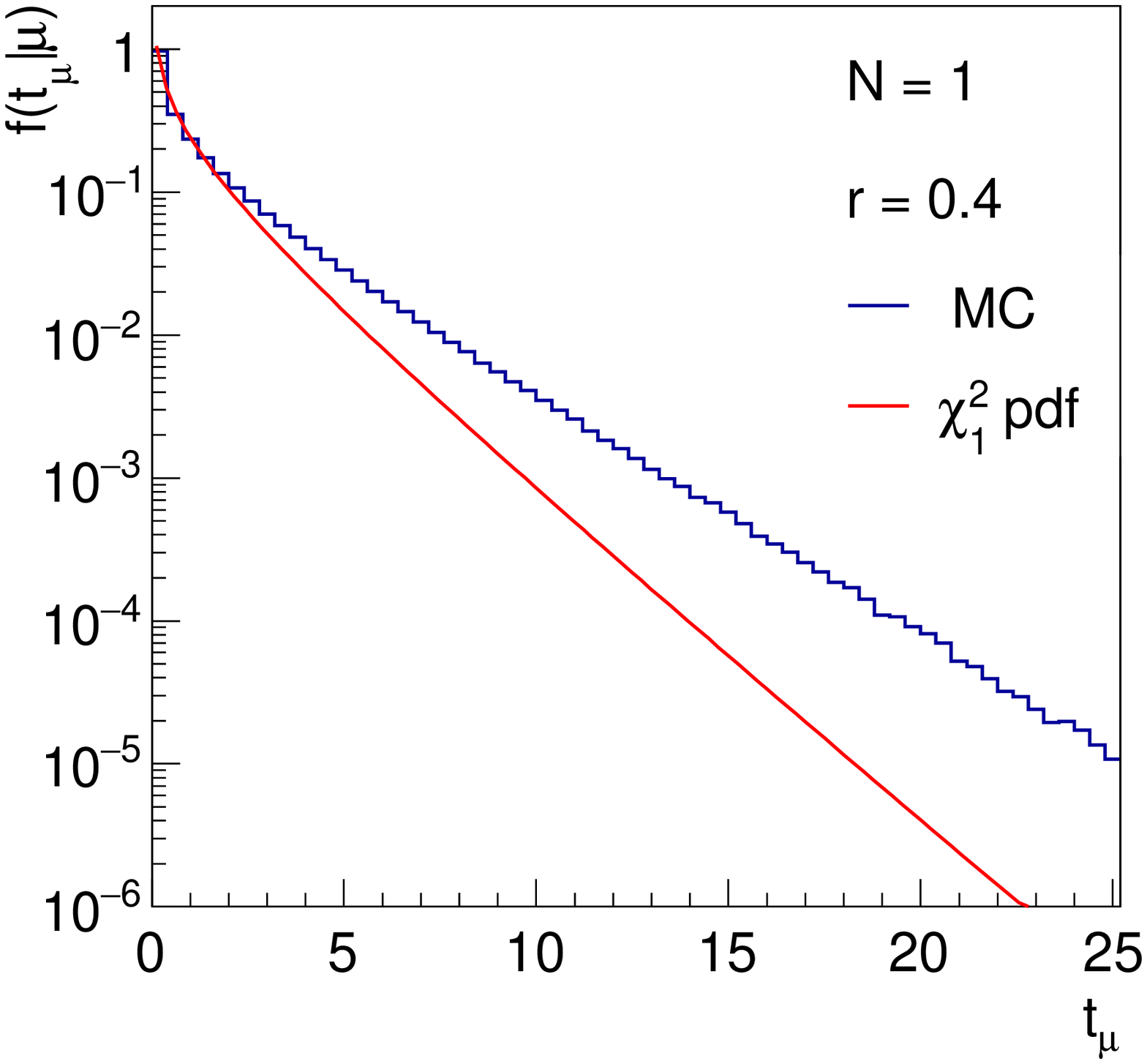}}
\put(10,0.0) {\includegraphics{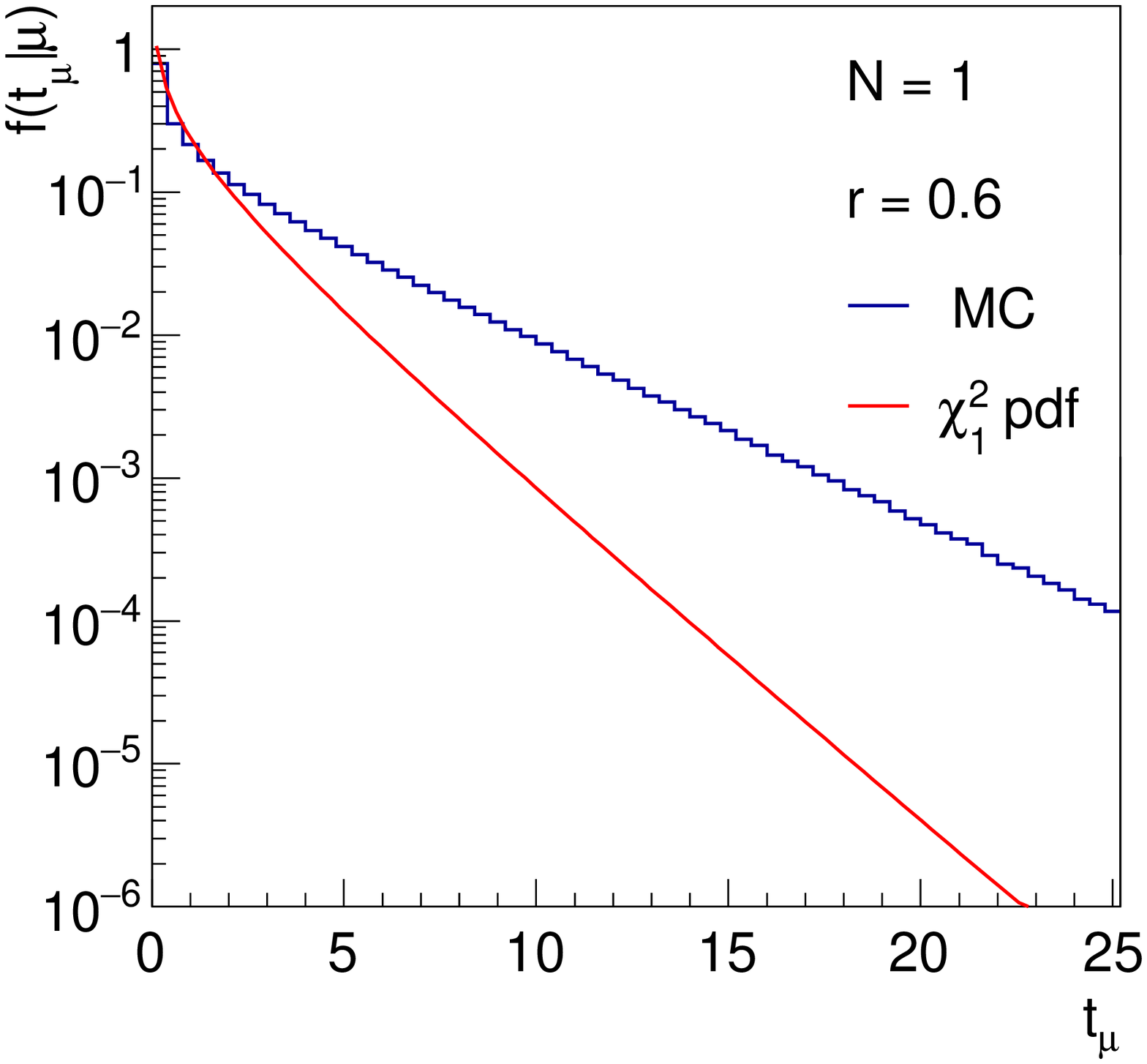}}
\put(7.8,9.5){(a)}
\put(15.8,9.5){(b)}
\put(7.8,4.5){(c)}
\put(15.8,4.5){(d)}
\end{picture}
\caption{\small Distributions of the test variable $t_{\mu}$ 
for a single Gaussian distributed measurement with relative
error-on-error $r$.}
\label{fig:tmudist}
\end{figure*}
\renewcommand{\baselinestretch}{1}
\small\normalsize

Depending on the size of the test being carried out or equivalently
the confidence level of the interval, one may find that the asymptotic
approximation is inadequate.  In such a case one may wish to use the
Monte Carlo simulation to determine the distribution of the test
statistic.  Alternatively one can modify the statistic so that its
distribution is better approximated by the asymptotic form, as
described in the following section.

\subsection{Bartlett correction for profile likelihood-ratio statistic}
\label{sec:bartlett}

The likelihood-ratio statistic can be modified so as to follow more
closely a chi-square distribution using a type of correction due to
Bartlett \cite{bib:Bartlett,bib:Cordeiro,bib:Brazzale}.  This method
has received some limited notice in Particle Physics
\cite{bib:Demortier} but has not been widely used in that field.  The
basic idea is to determine the mean value $E[t_{\mu}]$ of the original
statistic.  In the asymptotic limit, this should be equal to the
number of degrees of freedom $n_{\rm d}$ of the chi-square
distribution, which in this example is $n_{\rm d} = 1$.  One then
defines a modified statistic

\begin{equation}
t_{\mu}^{\prime} = \frac{n_{\rm d}}{E[t_{\mu}]} t_{\mu} \,,
\end{equation}

\noindent so that by construction $E[t^{\prime}_{\mu}] = n_{\rm d}$.
It was shown by Lawley \cite{bib:Lawley56} that the modified statistic
approaches the reference chi-squared distribution with a difference of
order $n^{-3/2}$, where here the effective sample size $n$ is related
to the parameter $r$ by $n = 1 + 1/2r^2$ (cf.\ Eqs.~(\ref{eq:alphan})
and (\ref{eq:alphai})).

One could in principle find the expectation value $E[t_{\mu}]$ by the
Monte Carlo method.  But for the method to be convenient to use one
would like to determine the Bartlett correction without resorting to
simulation.  By expanding the expectation value 

\begin{equation}
\label{eq:meantmudef}
E[t_{\mu}] = \int \int t_{\mu}(y, v) \, f(y,v | \mu, \sigma^2) \, dy \, dv
\end{equation}

\noindent as a Taylor series in $r$ one finds

\begin{equation}
\label{eq:meantmu}
E[t_{\mu}] = 1 + 3 r^2 + c r^4 \,,
\end{equation}

\noindent where the coefficient of the $r^4$ term is found numerically
to be $c \approx 2$ with an accuracy of around 10\%.  Dividing
$t_{\mu}/n_{\rm d}$ (here with $n_{\rm d} = 1$) from
Eq.~(\ref{eq:tmudef}) by $E[t_{\mu}]$ to obtain the Bartlett-corrected
statistic therefore gives

\begin{equation}
\label{eq:tmuprime}
t^{\prime}_{\mu} = \frac{1 + 2 r^2}{2r^2(1 + 3r^2 + 2r^4)}
\ln \left[ 1 + \frac{2r^2}{v} (y - \mu)^2 \right] \,.
\end{equation}

\noindent In more complex problems one may not have a simple
expression for the expectation value needed in the Bartlett correction
and calculation by Monte Carlo may be necessary.

Distributions of $t^{\prime}_{\mu}$ are shown in
Fig.~\ref{fig:tmudistbc} along with Monte Carlo distributions.  As can
be seen by comparing the uncorrected distributions from
Figs.~\ref{fig:tmudist} to those in Fig.~\ref{fig:tmudistbc}, the
Bartlett correction is clearly very effective, as is needed when the
parameter $r$ is large.

\setlength{\unitlength}{1.0 cm}
\renewcommand{\baselinestretch}{0.9}
\begin{figure*}[htbp]
\begin{picture}(10.0,10.5)
\put(2,5.2) {\includegraphics{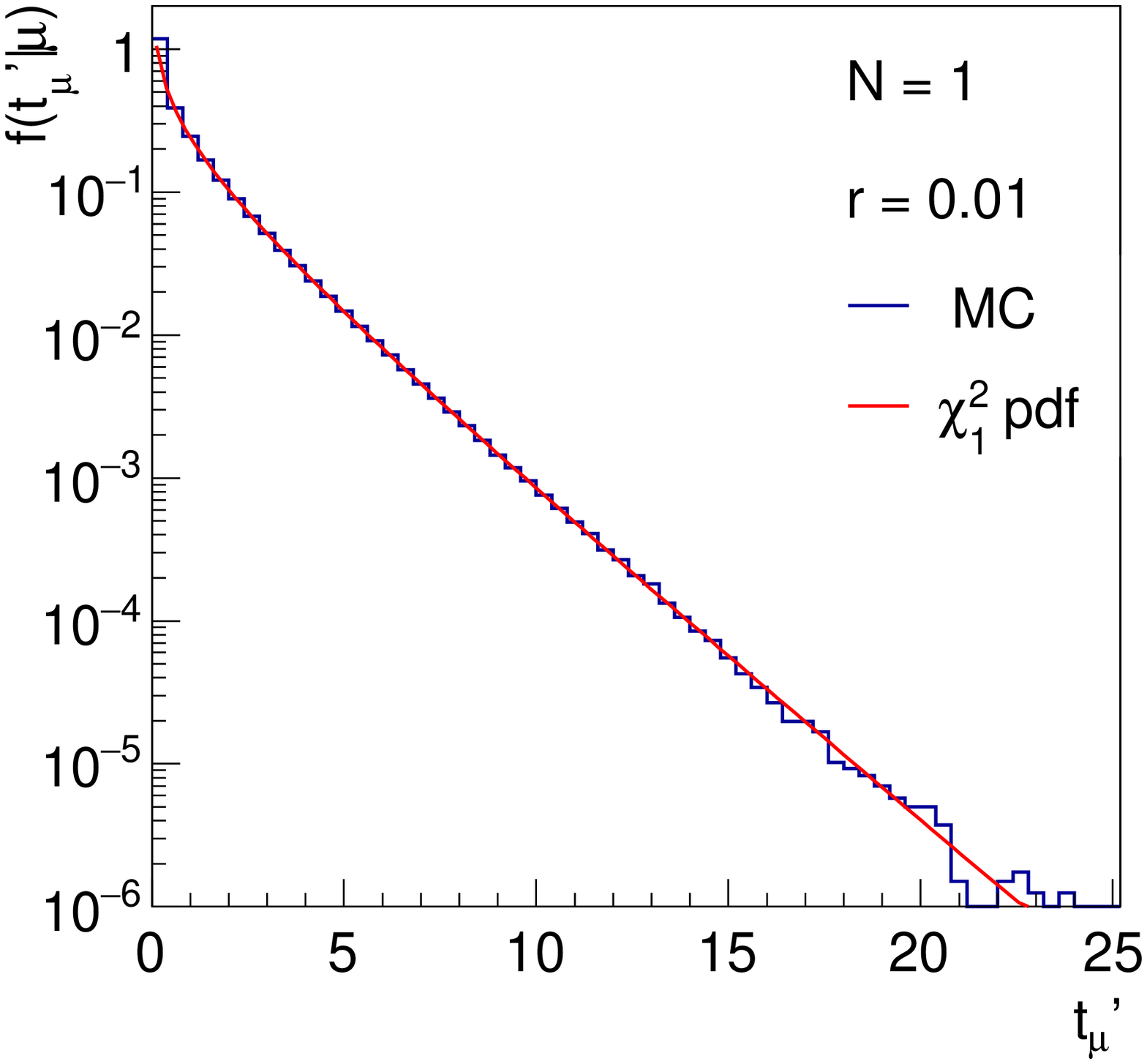}}
\put(10,5.2) {\includegraphics{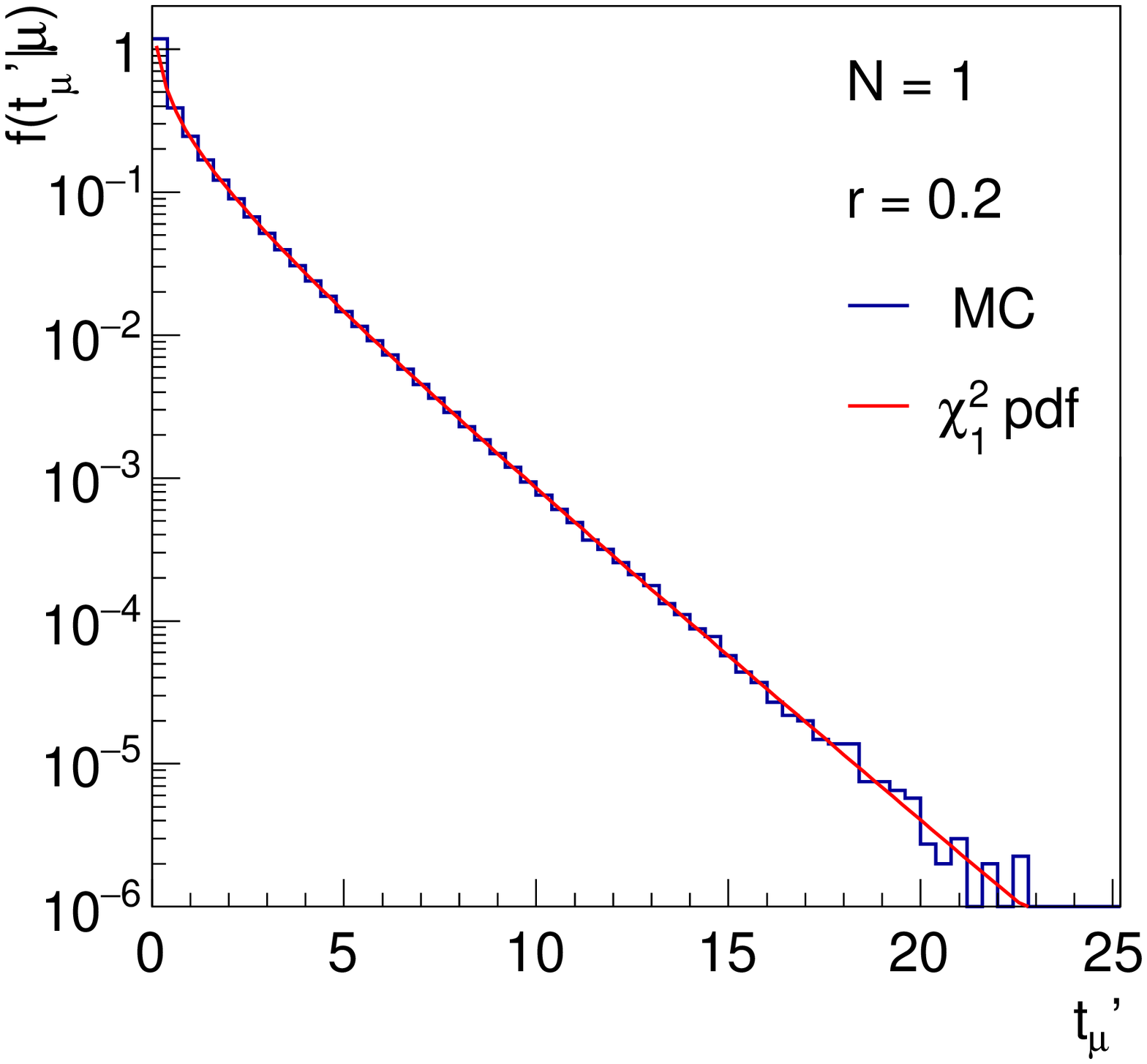}}
\put(2,0.0) {\includegraphics{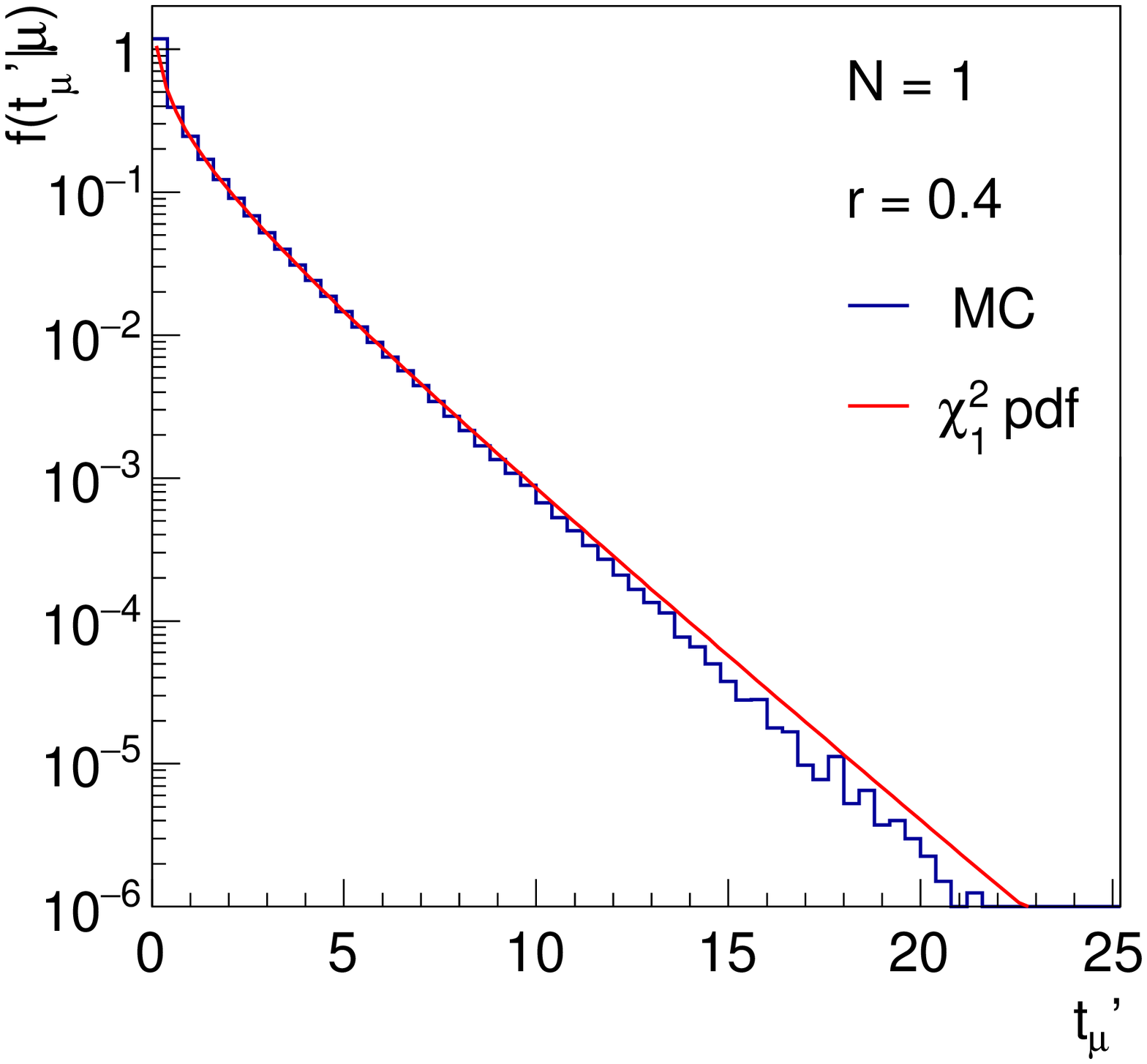}}
\put(10,0.0) {\includegraphics{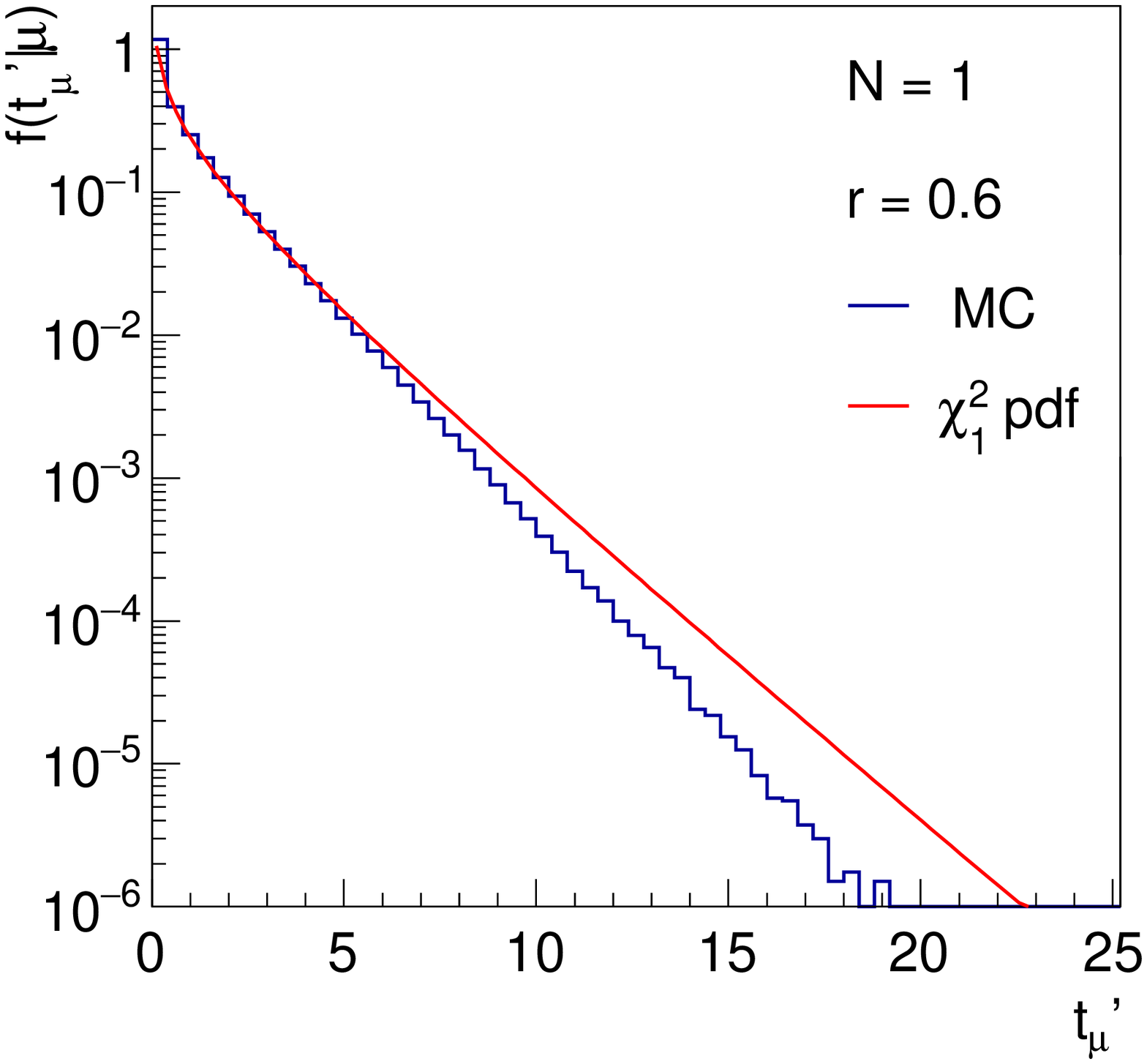}}
\put(7.8,9.5){(a)}
\put(15.8,9.5){(b)}
\put(7.8,4.5){(c)}
\put(15.8,4.5){(d)}
\end{picture}
\caption{\small  Distributions of the Bartlett-corrected 
test variable $t_{\mu}^{\prime}$ for a single Gaussian distributed 
measurement with relative error-on-error $r$.}
\label{fig:tmudistbc}
\end{figure*}
\renewcommand{\baselinestretch}{1}
\small\normalsize

\subsection{Confidence intervals for the single-measurement model}
\label{sec:simpleconf}

In the simple model explored in this section one can use the measured
values of $y$ and $v$ to construct a confidence interval for the
parameter of interest $\mu$.  The probability that the interval
includes the true value of $\mu$ (the coverage probability) can then
be studied as a function of the relative error on the error $r$.  What
emerges is that the interval based on the chi-squared distribution of
$t_{\mu}$ has a coverage probability substantially less than the
nominal confidence level, but that this can be greatly improved by use
of the Bartlett-corrected interval.

To derive exact confidence intervals for $\mu$ we can use the
fact that

\begin{equation}
\label{eq:xforsmm}
z = \frac{y - \mu}{\sqrt{v}}
\end{equation}

\noindent follows a Student's $t$ distribution for $\nu = 1/2r^2$
degrees of freedom (see, e.g., Ref.~\cite{Kendall1}).  From the
distribution of $z$ one can find the corresponding pdf of

\begin{equation}
\label{eq:tmux}
t_{\mu} = (1 + \nu) \ln \left[ 1 + \frac{z^2}{\nu} \right] \,,
\end{equation}

\noindent but in fact this is not directly needed.  Rather we can use
the pdf of $z$ to find confidence intervals for $\mu$ from the fact
that a critical region defined by $t_{\mu} > t_{\rm c}$ is equivalent
to the corresponding region of $z$ given by $z < -z_{\rm c}$ and $z >
z_{\rm c}$ where the boundaries of the critical regions in the two
variables are related by Eq.~(\ref{eq:tmux}).  Equivalently one can
say that the $p$-value of a hypothesized value of $\mu$ is the
probability, assuming $\mu$, to find $z$ further from zero than what
was observed, i.e.,

\begin{equation}
\label{eq:pmustudent}
p_{\mu} = 1 - \int_{-z_{\rm obs}}^{z_{\rm obs}} f(z) \, dz = 
2 \left( 1 - F\left( \frac{y - \mu}{\sqrt{v}};\nu \right) \right) \,,
\end{equation}

\noindent where $F(z;\nu)$ is the cumulative Student's $t$
distribution for $\nu = 1/2r^2$ degrees of freedom.

The boundaries of the confidence interval at confidence level
$\mbox{CL} = 1 - \alpha$ (here $\alpha$ refers to the size of the
statistical test, not the parameter $\alpha$ in the gamma
distribution) are found by setting $p_{\mu} = \alpha$ and solving for
$\mu$, which gives the upper and lower limits

\begin{equation}
\mu_{\pm} = y \pm \sqrt{v} z_{\alpha/2} \,.
\end{equation}

\noindent Here $z_{\alpha/2}$ is the $\alpha/2$ upper quantile of the
Student's $t$ distribution, i.e., the value of $z_{\rm obs}$ needed in
Eq.~(\ref{eq:pmustudent}) to have $p_{\mu} = \alpha$.

If one were to assume that the statistic $t_{\mu}$ follows the
asymptotic chi-squared distribution, then $z_{\alpha/2}$ is replaced
by

\begin{equation}
\label{eq:xa}
z_{\rm a} = \frac{1}{\sqrt{2} r} \left[
\exp \left( \frac{2 r^2 Q_{\alpha} }{1 + 2 r^2} \right) - 1 \right]^{1/2} \,.
\end{equation}

\noindent Here $Q_{\alpha} = F^{-1}_{\chi^2_1}(1 - \alpha)$ is
obtained from the quantile of the chi-squared distribution for one
degree of freedom.  And if the Bartlett-corrected statistic
$t^{\prime}_{\mu}$ is used to construct the interval, then the
$Q_{\alpha}$ in Eq.~(\ref{eq:xa}) is replaced by $Q_{\alpha}
E[t_{\mu}]$, where $E[t_{\mu}] = 1 + 3r^2 + 2r^4$ is the expectation
value of $t_{\mu}$ from Eq.~(\ref{eq:meantmu}).  The half-width of the
interval measured in units of the estimated standard deviation
$\sqrt{v}$, i.e., $z_{\alpha/2}$ or $z_{\rm a}$, are shown in
Fig.~\ref{fig:pcov}(a) as a function of the $r$ parameter.

The probability $P_{\rm c}$ for the confidence interval to cover the
true value of $\mu$ is by construction equal to $1 - \alpha$ for the
exact confidence interval.  For the interval based on the asymptotic
distribution of the test statistic this is

\begin{equation}
\label{eq:pcov}
P_{\rm c} = \int_{-z_{\rm a}}^{z_{\rm a}} f_{\chi^2_1}(z) \, dz
= 2 F_{\chi^2_1}(z_{\rm a}) - 1 \,,
\end{equation}

\noindent where $F_{\chi^2_1}$ is the cumulative chi-squared
distribution for one degree of freedom and $z_{\rm a}$ is given by
Eq.~(\ref{eq:xa}), with $Q_{\alpha}$ replaced by $Q_{\alpha}
E[t_{\mu}]$ for the Bartlett-corrected case.

The interval half-widths and coverage probabilities based on $t_{\mu}$
and $t^{\prime}_{\mu}$ are shown in Figs.~\ref{fig:pcov}.  As can be
seen, the interval based on the Bartlett-corrected statistic is very
close to the exact one, and its coverage is close to the nominal
$1-\alpha$ for relevant values of $r$.

\setlength{\unitlength}{1.0 cm}
\renewcommand{\baselinestretch}{0.9}
\begin{figure*}[htbp]
\begin{picture}(10.0,6.5)
\put(1,-0.2)
{\includegraphics{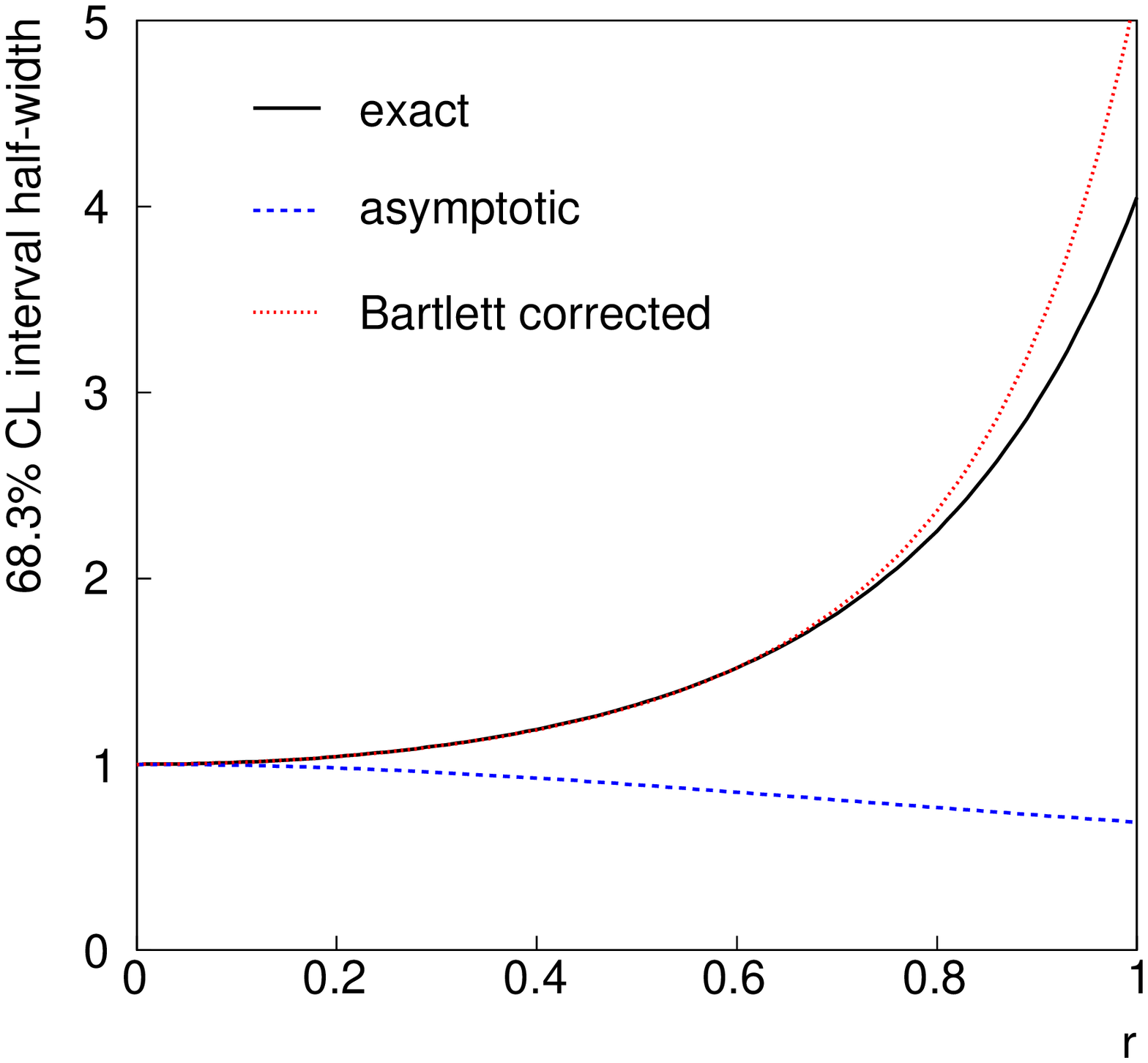}}
\put(10,-0.2)
{\includegraphics{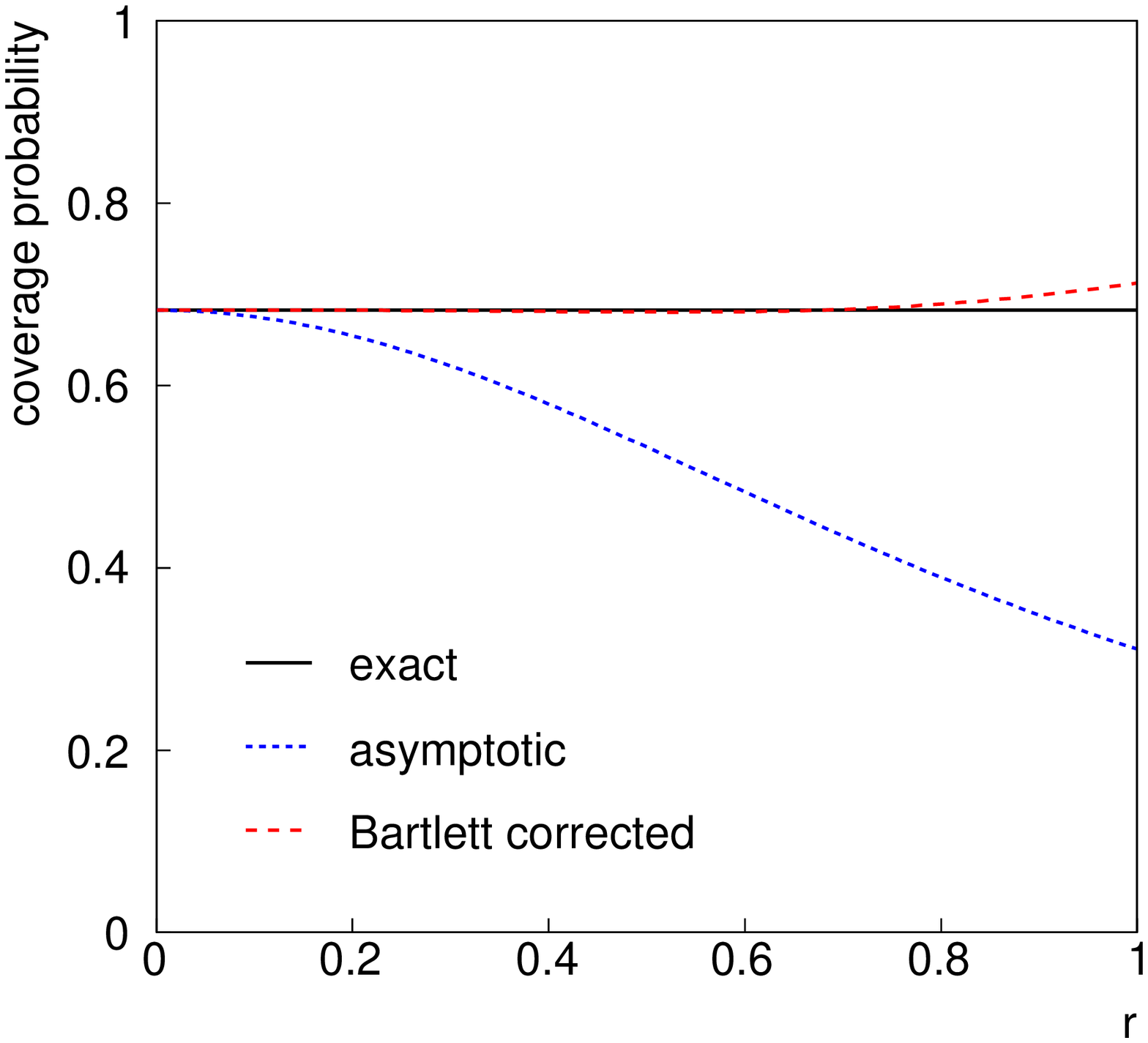}}
\put(8.,5.5){(a)}
\put(17,5.5){(b)}
\end{picture}
\caption{\small Plots of (a) the interval half-width in
units of the estimated standard deviation $\sqrt{v}$ and (b) coverage
  probability of the 68.3\% CL confidence intervals for $\mu$.}
\label{fig:pcov}
\end{figure*}
\renewcommand{\baselinestretch}{1}
\small\normalsize

As seen from the distributions in Figs.~\ref{fig:tmudist} and
\ref{fig:tmudistbc} for the single-measurement model, the agreement
with the asymptotic form worsens for increasing values of the test
statistic.  For $Z = \sqrt{t_{\mu}}$ of 4 (a four standard-deviation
significance; see, e.g., Ref.~\cite{asimov}), the Bartlett-corrected
statistic is close to the asymptotic form for $r = 0.2$, with a small
but visible departure for $r = 0.4$.  In contrast, for a 68.3\%
confidence level (corresponding to $\sqrt{t_{\mu}} = 1$), one sees
from Fig.~\ref{fig:pcov}(a) that the Bartlett corrected interval is in
satisfactory agreement with the exact interval out to $r \approx 1$.
For a more complicated analysis with multiple measurements having
different $r_i$ parameters one would need to check the validity of
asymptotic distributions with Monte Carlo.

\section{Least-squares fitting and averaging measurements}
\label{sec:ls}

An important application of the model described in
Sec.~\ref{sec:gammamod} is the least-squares fit of a curve, or as a
special case of this, the average of a set of measurements.  Suppose
the data consist of $N$ independent Gaussian distributed values $y_i$,
with mean and variance

\begin{eqnarray}
E[y_i] & =  & \varphi(x_i; \bvec{\mu}) + \theta_i \,, \\*[0.2 cm]
V[y_i] & = & \sigma^2_{y_i} \,.
\end{eqnarray}

\noindent Here the nuisance parameters $\theta_i$ represent a
potential bias or offset.  The function $\varphi(x_i;\bvec{\mu})$ plus
the bias $\theta_i$ gives the mean of $y_i$ as a function of a control
variable $x$, and it depends on a set of $M$ parameters of interest
$\bvec{\mu} = (\mu_1, \ldots, \mu_M)$.  That is, the probability
$P(\bvec{y} | \bvec{\varphi}, \bvec{\theta})$ in Eq.~(\ref{eq:pxu})
becomes

\begin{equation}
P(\bvec{y} | \bvec{\mu}, \bvec{\theta}) = \prod_{i=1}^N
\frac{1}{\sqrt{2 \pi} \sigma_{y_i}} 
e^{-(y_i - \varphi(x_i; \bvec{\mu}) - \theta_i)^2 / 2 \sigma_{y_i}^2} \,.
\end{equation}

As before suppose the nuisance parameters $\theta_i$ are constrained by
$N$ corresponding independent Gaussian measurements $u_i$, with mean
and variance

\begin{eqnarray}
E[u_i] & =  & \theta_i \,, \\*[0.2 cm]
V[u_i] & = & \sigma^2_{u_i} \,.
\end{eqnarray}

\noindent Often the best estimates of a potential bias $\theta_i$ will
be $u_i = 0$ for the actual measurement, but formally the $u_i$ are
treated as random variables that would fluctuate upon repetition of
the experiment.  Therefore the full log-likelihood or equivalently $-
2 \ln L(\bvec{\mu}, \bvec{\theta})$ is up to an additive constant given by

\begin{equation}
\label{eq:loglikelihood2}
-2 \ln L(\bvec{\mu}, \bvec{\theta}) = 
\sum_{i=1}^N \left[ \frac{(y_i - \varphi(x_i; \bvec{\mu}) - \theta_i)^2}
{\sigma_{y_i}^2}
+ \frac{(u_i - \theta_i)^2}{\sigma_{u_i}^2} \right] \,.
\end{equation}

\noindent That is, if we consider the $\sigma_{u_i}$ as known, then
maximum-likelihood estimators are obtained by the minimum of the sum
of squares (\ref{eq:loglikelihood2}) which is the usual formulation of
the method of least squares.

The next step will be to treat the $\sigma_{u_i}$ as adjustable
parameters but before doing this is it interesting to note that by
profiling over the nuisance parameters $\theta_i$, one finds the
profile likelihood

\begin{equation}
\label{eq:profloglikelihood2}
-2 \ln L^{\prime}(\bvec{\mu}) = 
\sum_{i=1}^N \frac{(y_i - \varphi(x_i; \bvec{\mu}) - u_i)^2}
{\sigma_{y_i}^2 + \sigma_{u_i}^2}  
\equiv \chi^2(\bvec{\mu}) \,.
\end{equation}

\noindent That is, the same result is obtained by using the usual
method of least squares with statistical and systematic uncertainties
added in quadrature.  This procedure gives the Best Linear Unbiased
Estimator (BLUE), which is widely used in Particle Physics,
particularly for the problem of averaging a set of measurements as
described in Refs.~\cite{aitken,lyons,valassi,bib:nisius2004}.

Returning to the full dependence on $\bvec{\mu}$ and $\bvec{\theta}$
and following the model of Sec.~\ref{sec:gammamod} we now regard the
systematic variances $\sigma_{u_i}^2$ as free parameters for which we
have independent gamma distributed estimates $v_i$, with parameters
$\alpha_i$ and $\beta_i$ set by $\sigma_{u_i}^2$ and $r_i$ according
to Eqs.~(\ref{eq:alphai}) and (\ref{eq:betai}).  The log-likelihood
profiled over the $\sigma_{u_i}^2$ is (cf.\ Eq.~(\ref{eq:proflikmub})),

\begin{eqnarray}
\label{eq:proflikls}
-2 \ln L^{\prime}(\bvec{\mu}, \bvec{\theta}) & = &  
\sum_{i=1}^N \left[ \frac{(y_i - \varphi(x_i; \bvec{\mu}) - \theta_i)^2}
{\sigma_{y_i}^2} \right.
 \\*[0.3 cm]
& + & \left.
\left( 1  + \frac{1}{2 r_i^2} \right) \ln \left(
1 + 2 r_i^2 \frac{(u_i - \theta_i)^2}{v_i} 
\right)  \right] \,. \nonumber
\end{eqnarray}

\noindent To find the required estimators we need to solve the system
of equations

\begin{eqnarray}
\label{eq:dlnldmu}
\frac{\partial \ln L^{\prime}}{\partial \mu_i} & = & 0 
\quad i = 1, \ldots,M \,, \\*[0.3 cm]
\label{eq:dlnldtheta}
\frac{\partial \ln L^{\prime}}{\partial \theta_i} & = & 0 \,, 
\quad i = 1,\ldots,N .
\end{eqnarray}

\noindent Equation~(\ref{eq:dlnldtheta}) results in

\begin{eqnarray}
\label{eq:mubrel2}
\theta_i^3  & + &  \left[ - 2 u_i - y_i + \varphi_i \right] \theta_i^2
\nonumber \\*[0.3 cm]
& + & \left[ \frac{v_i + (1 + 2 r_i^2) \sigma_{y_i}^2}{2 r_i^2} +
2 u_i (y_i - \varphi_i) + u_i^2 \right] \theta_i \nonumber \\*[0.3 cm]
& + & \left[ (\varphi_i - y_i) \left( \frac{v_i}{2 r_i^2} + u_i^2 \right) -
\frac{(1 + 2 r_i^2) \sigma_{y_i}^2 u_i}{2 r_i^2} \right] = 0 \,,
\nonumber \\*[0.3 cm]
& & \quad i = 1, \ldots, N \,,
\end{eqnarray}

\noindent where here $\varphi_i = \varphi(x_i; \bvec{\mu})$.
Simultaneously solving all $M+N$ equations for $\bvec{\mu}$ and the
$\bvec{\theta}$ gives their ML estimators.  Solving for the $\theta_i$
for fixed $\bvec{\mu}$, i.e., fixed $\varphi_i$, gives the profiled
values $\hat{\hat{\theta}}_i$.
Equations~(\ref{eq:mubrel2}) are cubic in $\theta_i$ and so can be
solved in closed form giving either one or three real roots.  In the
case of three roots, the one is chosen that maximizes $\ln
L^{\prime}$.

Using the profile log-likelihood from Eq.~(\ref{eq:proflikls}) one can
use, for example, the test statistic $t_{\mu}$ defined in
Eq.~(\ref{eq:tmudef}) to find confidence regions for $\bvec{\mu}$
following the general procedure outlined in Sec.~\ref{sec:confint}.
Examples of this will be shown in Sec.~\ref{sec:ave}.

\subsection{Goodness of fit}
\label{sec:gof}

In the usual method of least squares, the minimized sum of squares
$\chi^2_{\rm min} = \chi^2(\hat{\bvec{\mu}})$ based on
Eq.~(\ref{eq:profloglikelihood2}) is often used to quantify the
goodness-of-fit.  Because it is constructed as a sum of squares of
Gaussian distributed quantities, one can show (see, e.g.,
Ref.~\cite{Kendall2}) that its sampling distribution is chi-squared
for $N-M$ degrees of freedom, and the $p$-value of the hypothesis that
the true model lies somewhere in the parameter space of $\bvec{\mu}$ is
thus

\begin{equation}
\label{eq:pvalfromchi2}
p = \int_{\chi^2_{\rm min}}^{\infty} f_{\chi^2_{N-M}} (x) \, dx \,.
\end{equation}

When using the gamma error model presented above, the quantity $-2\ln
L^{\prime}(\bvec{\mu}, \bvec{\theta})$ is no longer a simple sum of
squares.  Nevertheless one can construct the statistic that will play
the same role as the minimized $\chi^2(\bvec{\mu})$ by considering the
model in which the means $\varphi(x_i, \bvec{\mu})$, which depend on
the $M$ parameters of interest $\bvec{\mu}$, are replaced by a vector
of $N$ independent mean values, one for each of the measurements:
$\bvec{\varphi} = (\varphi_1, \ldots, \varphi_N)$.  By requiring that
the $\varphi_i$ are given by $\varphi(x_i, \bvec{\mu})$ one imposes
$N-M$ constraints and restricts the more general hypothesis to an
$M$-dimensional subspace.  One can then construct the likelihood ratio
statistic

\begin{equation}
\label{eq:qstat}
q = - 2 \ln \frac
{L^{\prime}(\hat{\bvec{\mu}}, \hat{\hat{\bvec{\theta}}})}
{L^{\prime}(\hat{\bvec{\varphi}}, \hat{\bvec{\theta}})} \,,
\end{equation}

\noindent where the numerator contains the $M$ fitted parameters of
interest $\hat{\bvec{\mu}}$, and in the denominator one fits all
$N$ of the $\varphi_i$.

When fitting separate values of $\varphi_i$ and $\theta_i$ for each
measurement (the ``saturated model''), one can see from inspection
that the maximized value of $\ln L^{\prime}(\bvec{\varphi},
\bvec{\theta})$ is zero, and therefore the statistic $q$ becomes

\begin{eqnarray}
\label{eq:qstat2}
q & = & \min_{\bvec{\mu}, \bvec{\theta}}  \; \sum_{i=1}^N \left[ 
\frac{(y_i - \varphi(x_i; \bvec{\mu}) - \theta_i)^2}{\sigma_{y_i}^2} 
\right. \nonumber \\*[0.3 cm]
& + &  \left. \left( 1  + \frac{1}{2 r_i^2} \right) \ln \left(
1 + 2 r_i^2 \frac{(u_i - \theta_i)^2}{v_i} \right)  \right] 
\end{eqnarray}

\noindent According to Wilks' theorem \cite{bib:Wilks}, in the limit
where the estimators $\hat{\bvec{\mu}}$ and $\hat{\bvec{\theta}}$ are
Gaussian distributed, $q$ will follow a chi-squared pdf for $N-M$
degrees of freedom.  The statistic $q$ thus plays the same role as the
minimized sum of squares $\chi^2_{\rm min}$ in the usual method of
least squares.  In the case of Eq.~(\ref{eq:qstat2}), however, the
chi-squared approximation is not exact.  One can see this from the
fact that the $v_i$ are gamma rather than Gaussian distributed; the
Gaussian approximation holds only in the limit where the $r_i$ are
sufficiently small.

If all $r_i \rightarrow 0$, i.e., there is no uncertainty in the
reported systematic errors, then the statistic $q$ reduces to the
minimized sum of squares from the method of least squares or BLUE,
namely,

\begin{equation}
\label{eq:qstatls}
q = \sum_{i=1}^N \frac{(y_i - \mu(\hat{\bvec{\varphi}}))^2}{\sigma_{y_i}^2 
+ \sigma_{u_i}^2} \,.
\end{equation}

One can check in an example that the sampling distribution of $q$
follows a chi-squared distribution by generating measured values
$y_i$, $u_i$, and $s_i$ according to the model described in
Sec.~\ref{sec:proflike} using the following parameter values:
$\varphi_i = \mu = 10$, $\sigma_{y_i} = 1$, $\sigma_{u_i} = 1$ for all
$i = 1, \ldots, N$.  That is, the measurements are assumed to have the
same mean $\mu$ and the goal is to fit this parameter.  The resulting
distributions of $q$ are shown in Figs.~\ref{fig:qdist}(a) and (b) for
$N=2$ and $N=5$ using $r_i = 0.2$ for all measurements.  Overlayed on
the histograms is the chi-squared pdf for $N-1$ degrees of freedom.
Although the agreement is reasonably good there is still a noticeable
departure from the asymptotic distribution in the tails.  The same set
of curves is shown in Figs.~\ref{fig:qdist}(c) and (d) for $r_i =
0.4$, for which one sees an even greater discrepancy between the true
(i.e., simulated) and asymptotic distributions.

\setlength{\unitlength}{1.0 cm}
\renewcommand{\baselinestretch}{0.9}
\begin{figure*}[htbp]
\begin{picture}(10.0,10.5)
\put(2,5.2) {\includegraphics{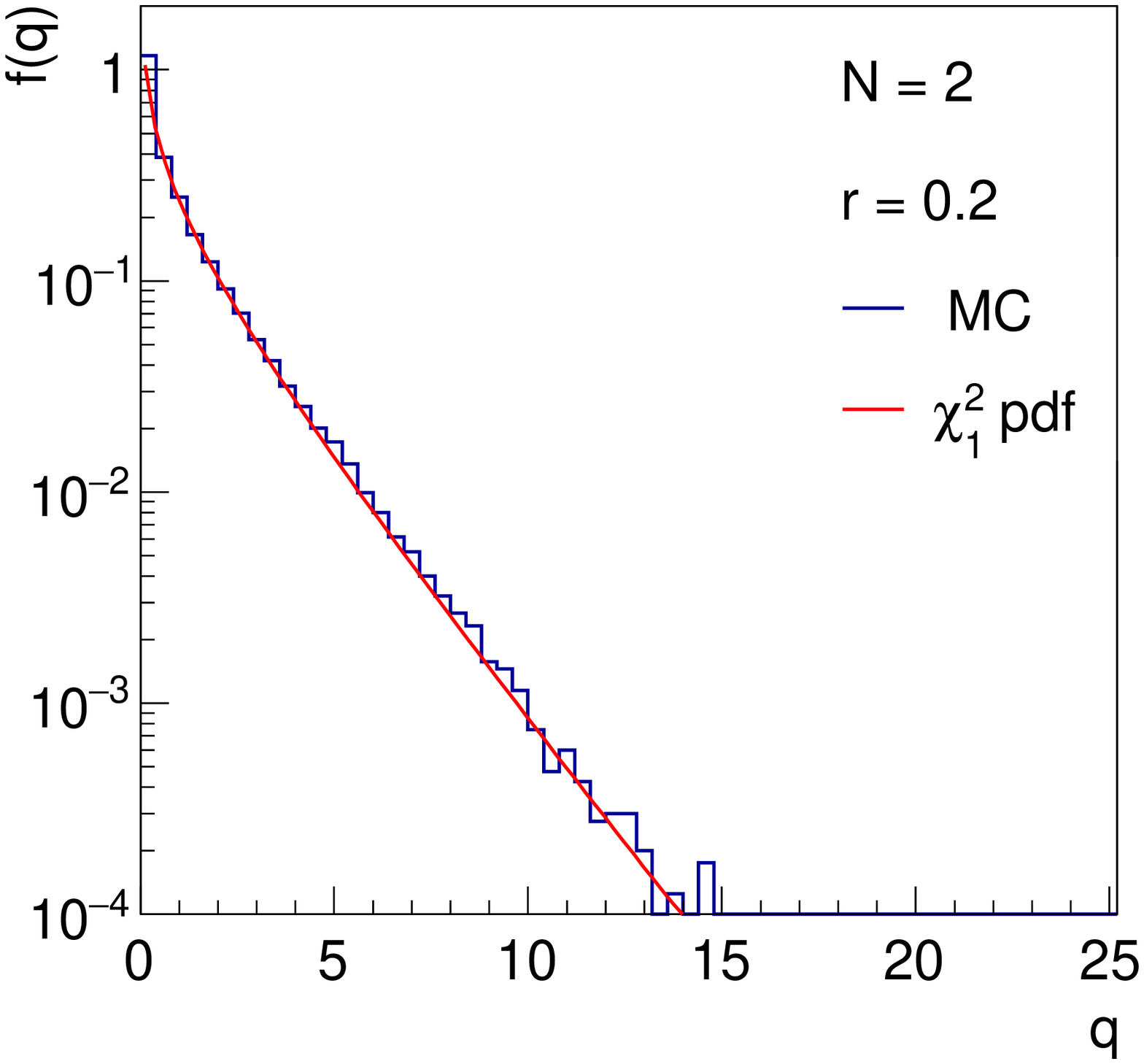}}
\put(10,5.2) {\includegraphics{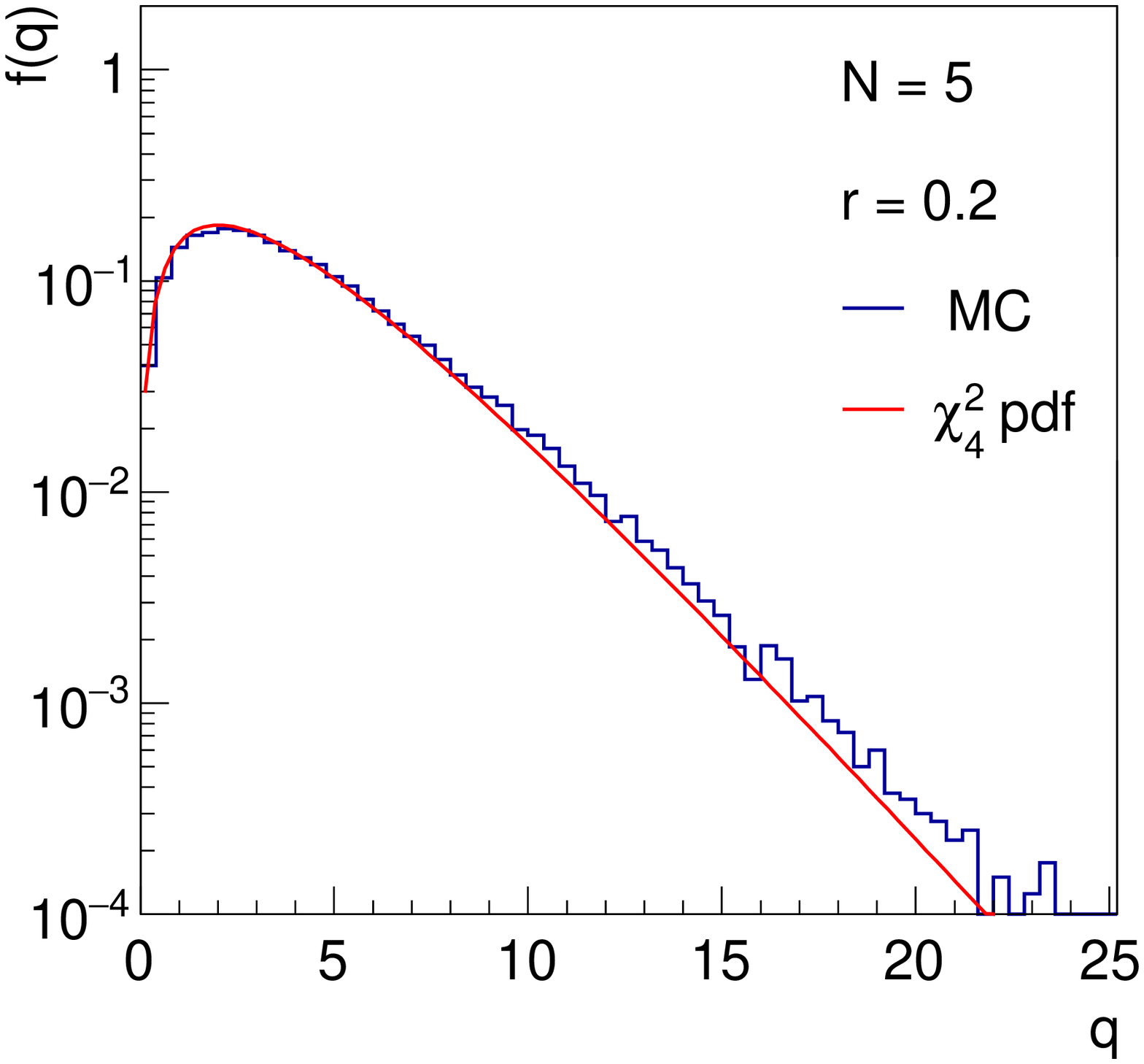}}
\put(2,0.0) {\includegraphics{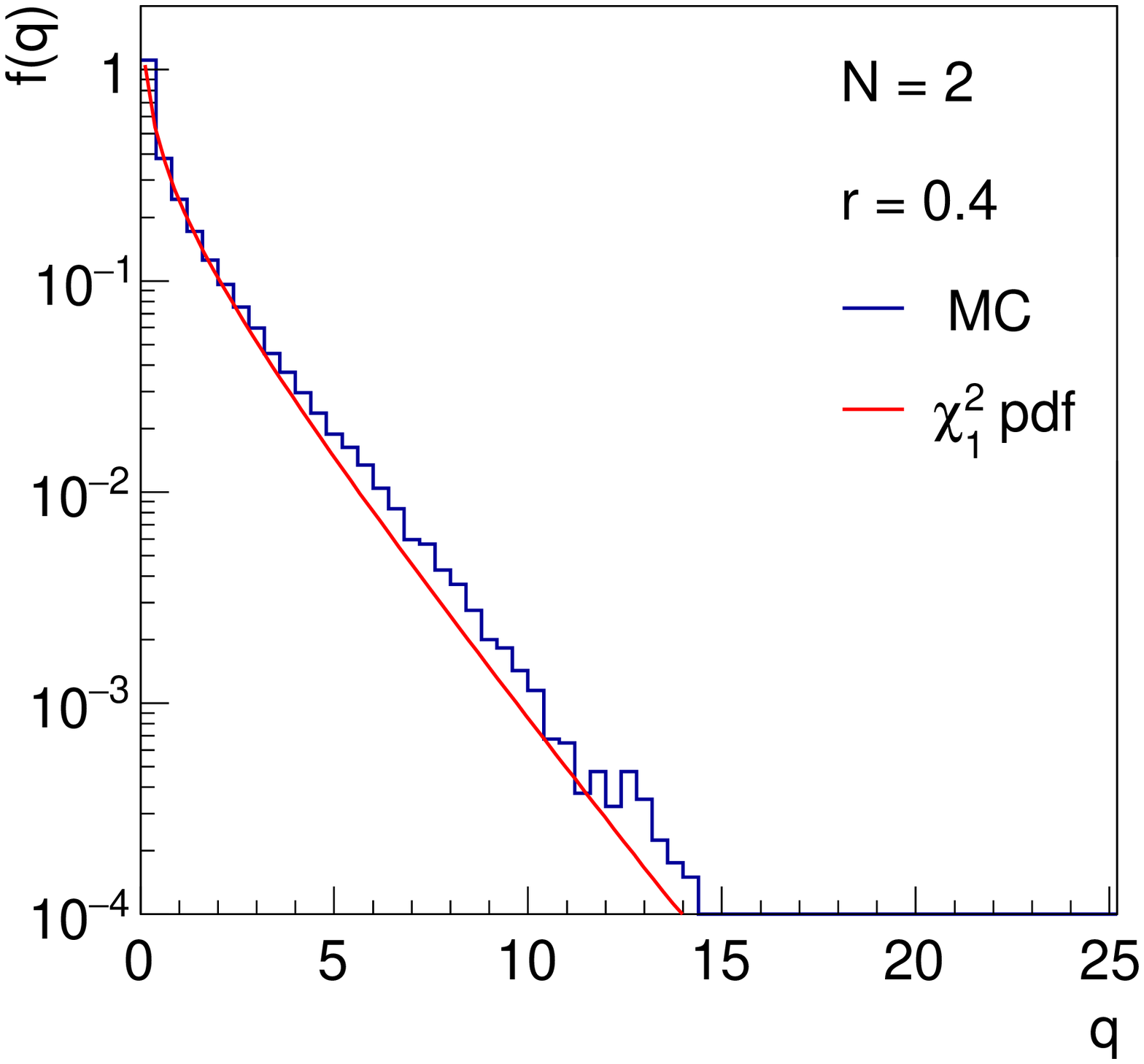}}
\put(10,0.0) {\includegraphics{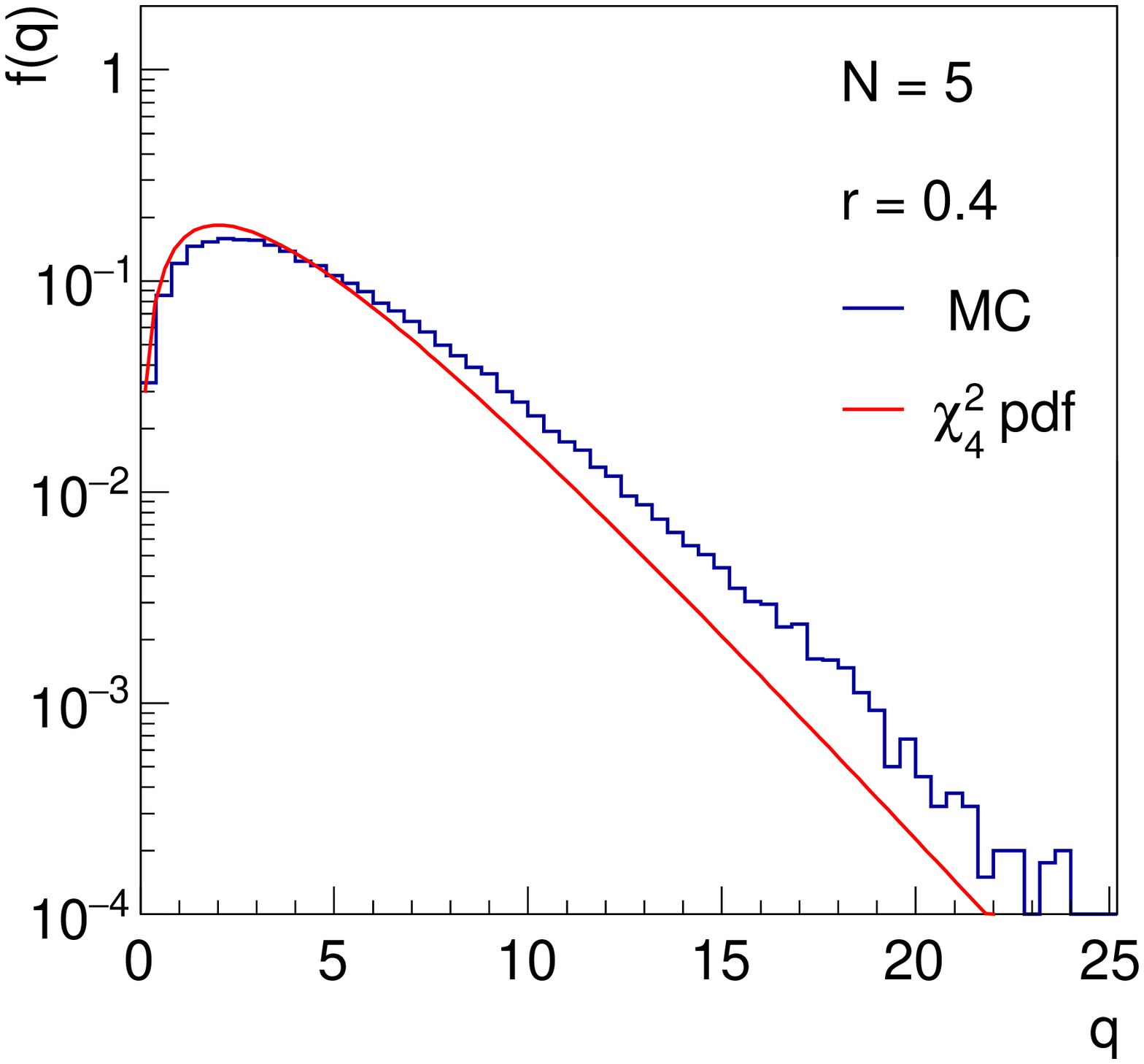}}
\put(7.8,9.5){(a)}
\put(15.8,9.5){(b)}
\put(7.8,4.5){(c)}
\put(15.8,4.5){(d)}
\end{picture}
\caption{\small Distributions of the test variable $q$ 
for averages of $N=2$ and $5$ values using $r = 0.2$ and $r=0.4$.}
\label{fig:qdist}
\end{figure*}
\renewcommand{\baselinestretch}{1}
\small\normalsize

One might need a $p$-value with an accuracy such that assumption of
the asymptotic distribution of $q$ is not adequate.  In such a case
one can use Monte Carlo to determine the correct sampling distribution
of $q$.  Alternatively, following the procedure of
Sec.~\ref{sec:bartlett} one can define a Bartlett-corrected statistic
$q^{\prime}$ as

\begin{equation}
\label{eq:qprime}
q^{\prime} = \frac{N-M}{E[q]} q \,,
\end{equation}

\noindent so that by construction $E[q^{\prime}] = N-M$ (in the
example above for a single fitted parameter $M = 1$).  Distributions
of $q^{\prime}$ corresponding to Fig.~\ref{fig:qdist} are shown in
Fig.~\ref{fig:qdistbc}, where the mean value $E[q]$ was itself found
from Monte Carlo simulation.  While one sees that the distributions of
$q^{\prime}$ are in better agreement with the Monte Carlo, visible
discrepancies remain.  And since here simulation was required to
determine the Bartlett correction, one could use it as well to find
the $p$-value directly.  The Bartlett correction is nevertheless
useful in such a situation because the number of simulated values of
$q$ required to estimate accurately $E[q]$ may be much less than what
one needs to find the upper tail area for a very high observed value
of the test statistic.

\setlength{\unitlength}{1.0 cm}
\renewcommand{\baselinestretch}{0.9}
\begin{figure*}[htbp]
\begin{picture}(10.0,10.5)
\put(2,5.2) {\includegraphics{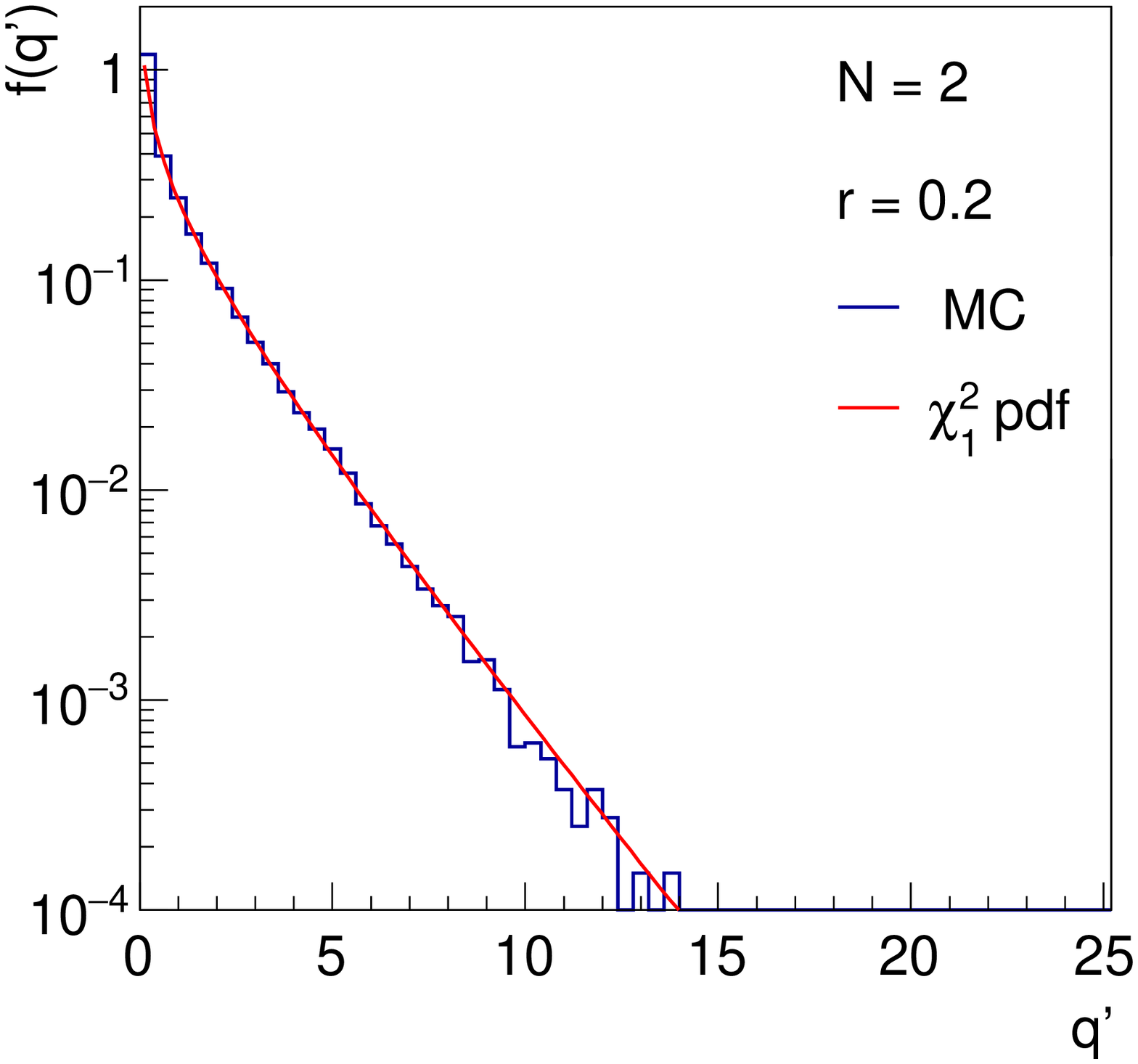}}
\put(10,5.2) {\includegraphics{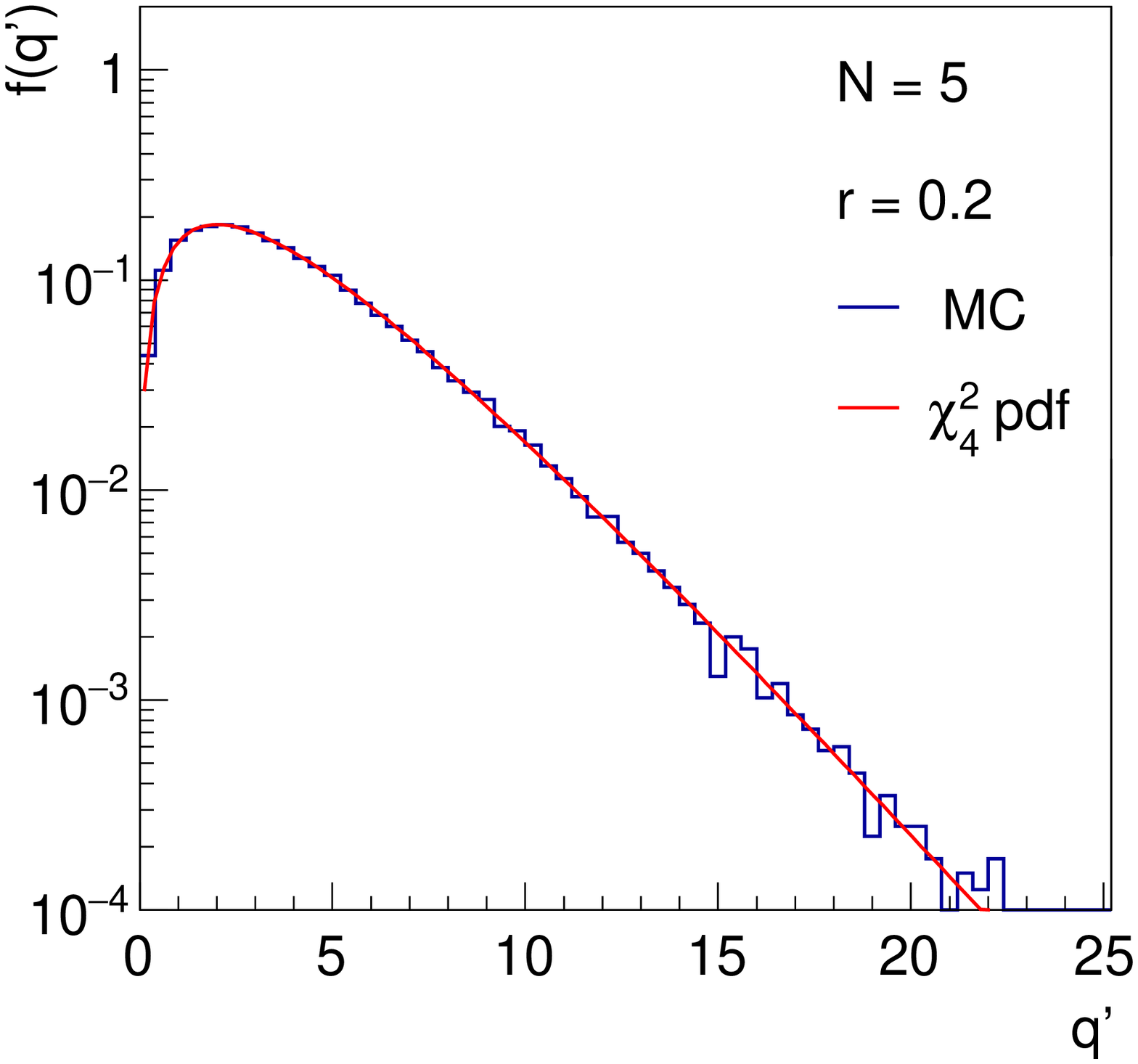}}
\put(2,0.0) {\includegraphics{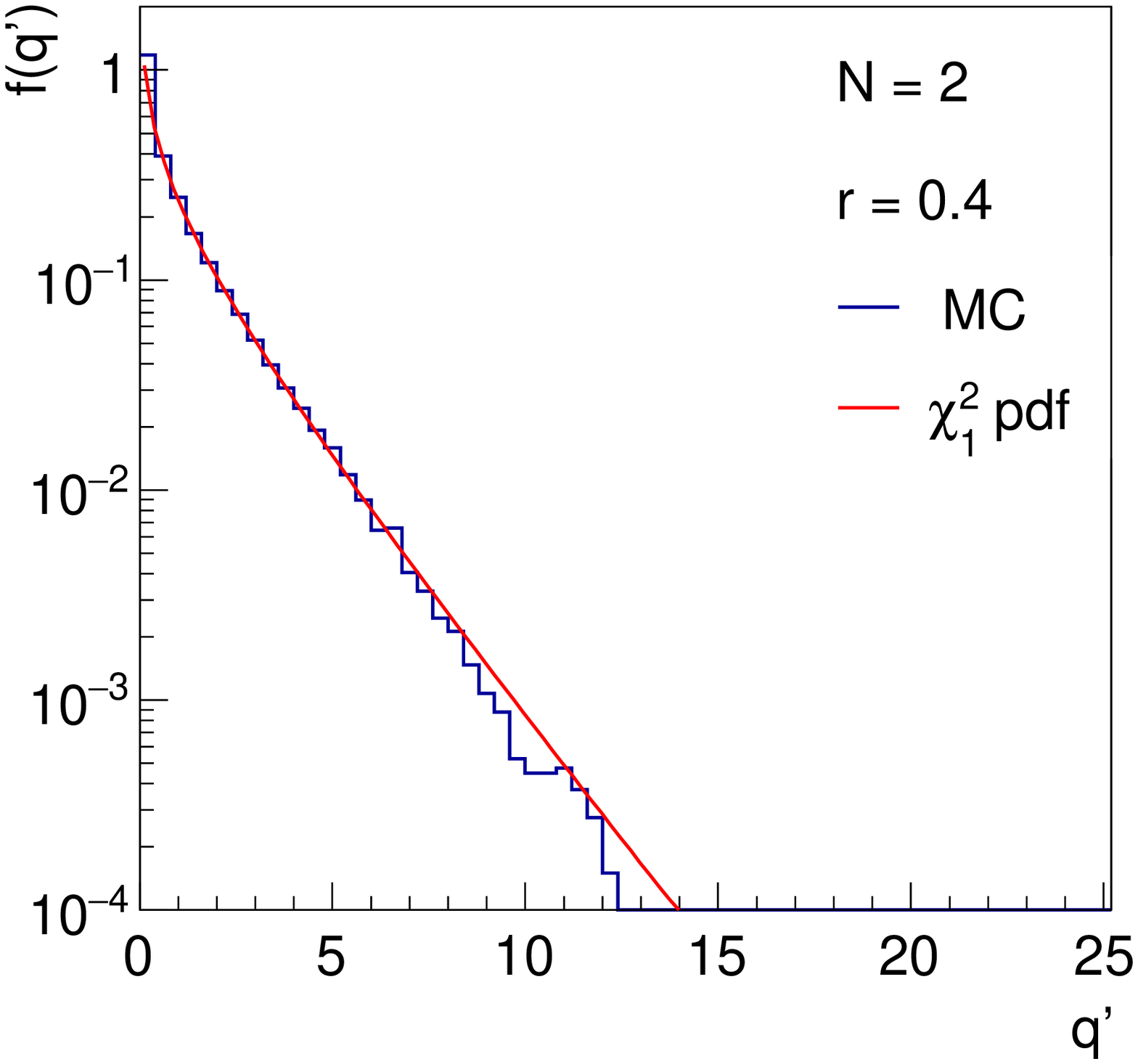}}
\put(10,0.0) {\includegraphics{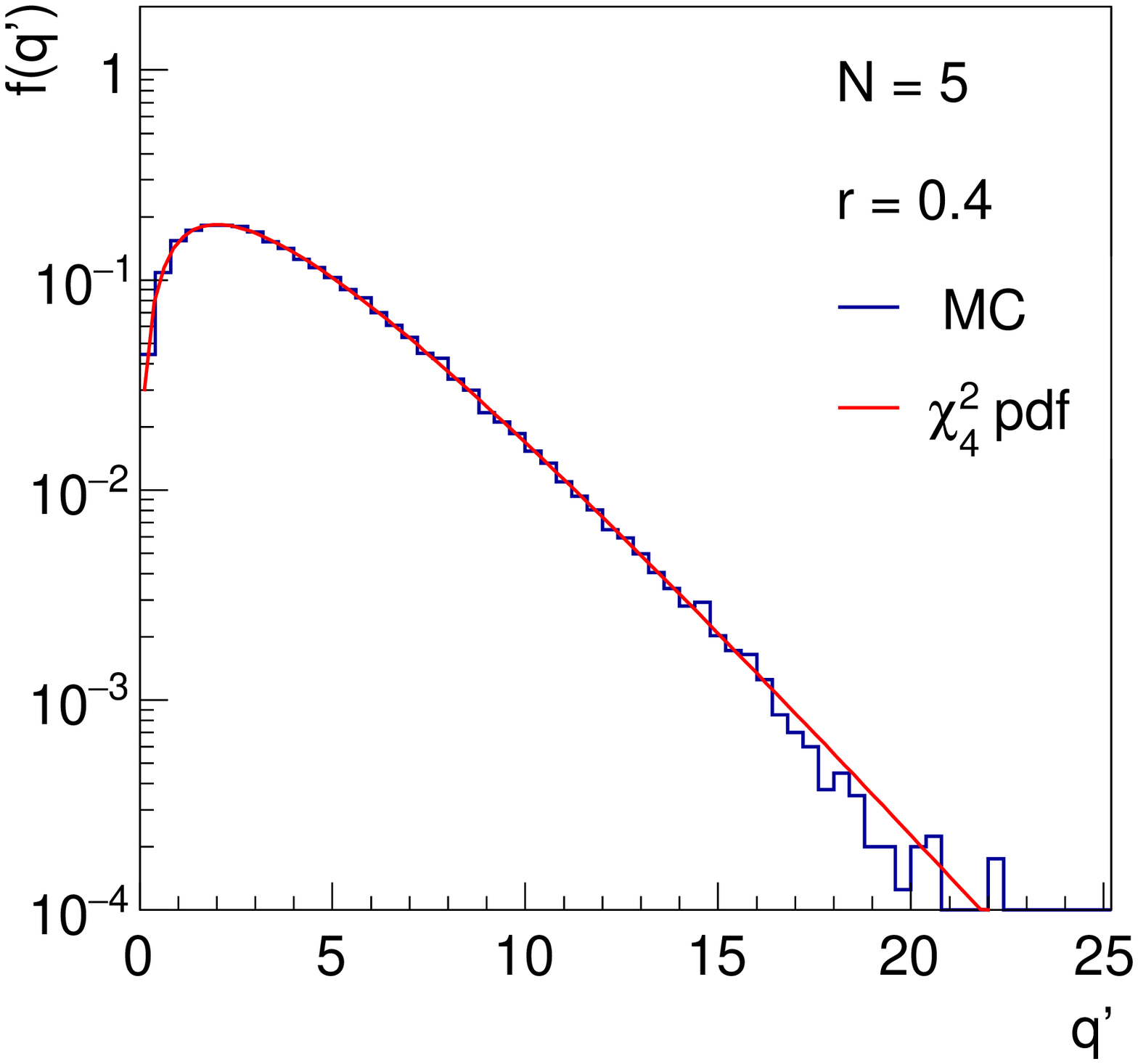}}
\put(7.8,9.5){(a)}
\put(15.8,9.5){(b)}
\put(7.8,4.5){(c)}
\put(15.8,4.5){(d)}
\end{picture}
\caption{\small Distributions of the Bartlett-corrected test variable 
$q^{\prime}$  for averages of $N=2$ and $5$ values using $r = 0.2$ and $r=0.4$.}
\label{fig:qdistbc}
\end{figure*}
\renewcommand{\baselinestretch}{1}
\small\normalsize

\subsection{Averaging measurements}
\label{sec:ave}

An important special case of a least-squares fit is the average of $N$
independent measurements, $y_1, \ldots, y_N$, of the same
quantity, 
i.e., the fit function $\varphi(x; \mu) = \mu$ is
in effect a horizontal line and the control variable $x$ does not
enter.  The expectation values of the measurements are thus

\begin{equation}
E[y_i] = \mu + \theta_i \,, \quad \quad  i = 1,\ldots,N\,,
\end{equation}

\noindent where the parameter of interest $\mu$ represents the desired
mean value and as before $\theta_i$ are the bias parameters.  As there
is one parameter of interest, the statistic $q$ follows asymptotically
a chi-squared distribution for $N$ degrees of freedom, although as we
have seen above this approximation breaks down as the $r_i$ increase.

As an example, consider the average of two independent measurements,
nominally reported as $y_i \pm \sigma_{y_i} \pm s_i$ for $i = 1, 2$,
in which the $\sigma_{y_i}$ represent the statistical uncertainties
and $s_i$ are the estimated systematic errors.  Suppose here these are
$\sigma_{y_i} = 1$ and $s_i = 1$ for both measurements, and that the
analyst reports values $r_i$ representing the relative accuracy of the
estimates of the systematic errors, which in this example we will take
to be equal to a common value $r$.  Furthermore suppose that the
observed values of $y_1$ and $y_2$ are $10 + \delta$ and $10 -
\delta$, respectively, and we will allow $\delta$ to vary.  For the
values of $\sigma_{y_i}$ and $s_i$ chosen in this example, the value
of $\delta$ corresponds to the significance of the discrepancy between
$y_1$ and $y_2$ in standard deviations under assumption of $r=0$.

Using the input values described above, the mean $\mu$, bias
parameters $\theta_i$, and systematic errors $\sigma_{u_i}$ are
adjusted to maximize the log-likelihood from Eq.~(\ref{eq:lnLfull}).
Figures~\ref{fig:intHalfWidth} show the half-width of the 68.3\%
confidence interval for $\mu$ as a function of the parameter $r$ for
different levels of $\delta$.  This interval corresponds to the
standard deviation $\sigma_{\hat{\mu}}$ when the $r_i$ are all small,
where the problem is the same as in least squares or BLUE.

In Fig.~\ref{fig:intHalfWidth}(a), the interval is based on
Eq.~(\ref{eq:confintbound}), i.e., it is determined by the point where
the profile log-likelihood drops by a fixed amount from its maximum
(in Particle Physics often referred to as the ``MINOS'' interval
\cite{bib:minuit}).  In Fig.~\ref{fig:intHalfWidth}(b), the interval
is found by solving for the value of $\mu$ where its $p$-value is
$p_{\mu} = \alpha$, and here $\alpha = 1 - 0.683 = 0.317$.  The
$p$-value depends, however, on the assumed values of the nuisance
parameters.  Here we use the values of $\theta_i$ and $\sigma^2_{u_i}$
profiled at the value of $\mu$ tested.  This technique is often called
``profile construction'' in Particle Physics \cite{bib:CranmerPC},
where it is widely used, and elsewhere called ``hybrid resampling''
\cite{bib:chuang,bib:sen}.  The resulting confidence interval will
have the correct coverage probability of $1 - \alpha$ if the nuisance
parameters are equal to their profiled values; elsewhere the interval
could under- or over-cover.  Although the intervals from profile
construction differ somewhat from those found directly on the
log-likelihood, they have the same qualitative behaviour.

\setlength{\unitlength}{1.0 cm}
\renewcommand{\baselinestretch}{0.9}
\begin{figure*}[htbp]
\begin{picture}(10.0,6.5)
\put(1,-0.2)
{\includegraphics{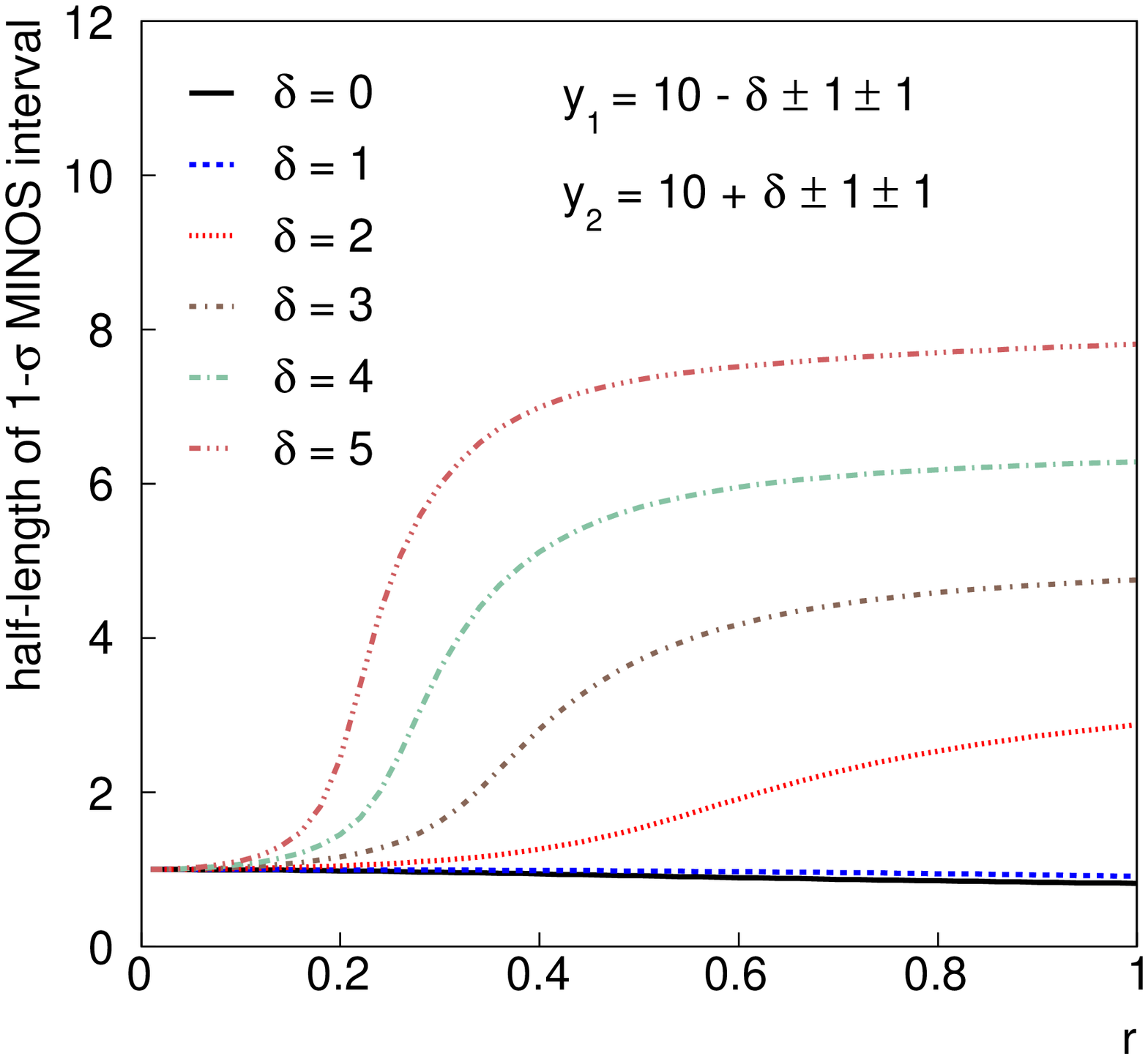}}
\put(10,-0.2)
{\includegraphics{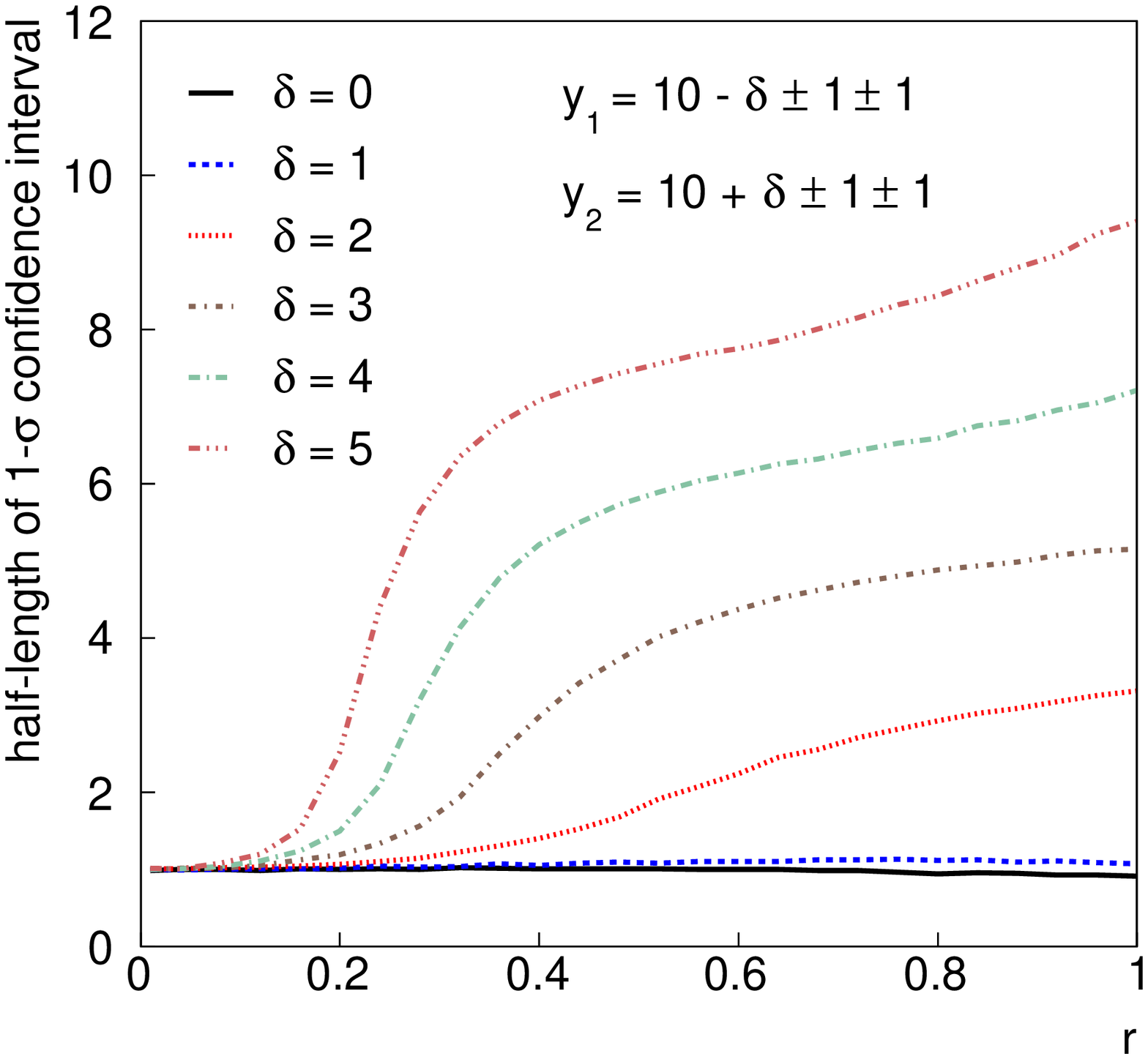}}
\put(8.,5.5){(a)}
\put(17,5.5){(b)}
\end{picture}
\caption{\small Plots of the half-length of the
1-$\sigma$ (68.3\%) central confidence interval for 
the parameter $\mu$ as a function of the relative uncertainty
on the systematic errors $r$ for different levels of discrepancy
$\delta$ between two averaged measurements.  Intervals
are derived (a) from the log-likelihood and (b) 
using ``profile construction'' (see text).} 
\label{fig:intHalfWidth}
\end{figure*}
\renewcommand{\baselinestretch}{1}
\small\normalsize

From Fig.~\ref{fig:intHalfWidth} one can extract several interesting
features.  First, if $r$ is small, that is, the systematic errors
$\sigma_{u_i}$ are very close to their estimated values $s_i$, then
the interval's half-length is very close to the standard deviation of
the estimator, $\sigma_{\hat{\mu}} = 1$, regardless of the level of
discrepancy between the two measured values.

Further, the effect of larger values of $r$ is seen to depend very
much on the level of discrepancy between the measured values.  If
$y_1$ or $y_2$ are very close (e.g., $\delta = 0$ or $1$), then the
length of the confidence interval can even be reduced relative to the
case of $r=0$.  If the measurements are in agreement at a level that
is better than expected, given the reported statistical and systematic
uncertainties, then one finds that the likelihood is maximized for
values of the systematic errors $\sigma_{u_i}$ that are smaller than
the initially estimated $s_i$.  And as a consequence, the confidence
interval for $\mu$ shrinks.

Finally, one can see that if the data are increasingly inconsistent,
e.g., in Fig.~\ref{fig:intHalfWidth} for $\delta \ge 4$, then the
effect of allowing higher $r$ is to increase the length of the
interval.  This is also a natural consequence of the assumed model,
whereby an observed level of heterogeneity greater than what was
initially estimated results in maximizing the likelihood for larger
values of $\sigma_{u_i}$ and consequently an increased confidence
interval size.  

The coverage properties of the intervals for the average of two
measurements example are investigated by generating data values $y_i$
for $i = 1,2$ according to a Gaussian with a common mean $\mu$ (here
10) and the standard deviations both $\sigma_{y_i} = 1$, and the $u_i$
are generated according to a Gaussian distributed with mean of
$\theta_i = 0$ and standard deviation $\sigma_{u_i} = 1$.  The values
$v_i$ are gamma distributed with parameters $\alpha_i$ and $\beta_i$
given by Eqs.~(\ref{eq:alphai}) and (\ref{eq:betai}) so as to
correspond $\sigma_{u_i} = 1$ and for different values of the
parameters $r_i$, taken here to be the same for both measurements.

Figure~\ref{fig:pcov_vs_r} shows the coverage probability for the
interval with nominal confidence level 68.3\% based on the
log-likelihood (the MINOS interval) and also using profile
construction (hybrid resampling), as a function of the $r$ parameter.
As seen in the figure, the coverage probability approximates the
nominal value reasonably well out to $r = 0.5$, where one finds
$P_{\rm cov} = 0.631$ and $0.667$ for MINOS and profile construction
respectively; at $r = 1$, the corresponding values are $0.564$ and
$0.617$ (the Monte Carlo statistical errors for all values is around
0.005).  Thus reasonable agreement is found with both methods but one
should be aware that the coverage probability may depart from the
nominal value for large values of $r$.

\setlength{\unitlength}{1.0 cm}
\renewcommand{\baselinestretch}{0.9}
\begin{figure}[htbp]
\begin{picture}(6.0,6.0)
\put(0.5,-0.5){\includegraphics{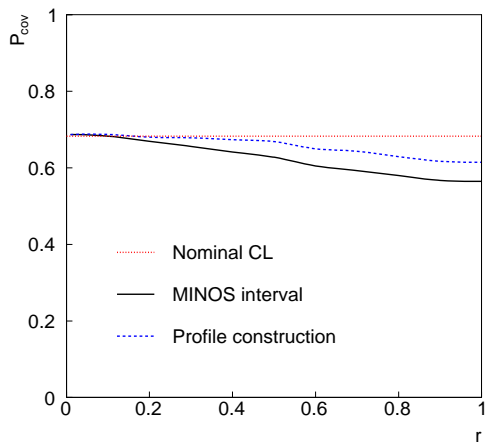}}
\end{picture}
\caption{\small  The coverage probability of
the intervals based on the likelihood (MINOS method)
and on profile construction (hybrid resampling) as a 
function of the parameter $r$ (see text).}
\label{fig:pcov_vs_r}
\end{figure}
\renewcommand{\baselinestretch}{1}
\small\normalsize

\subsection{Sensitivity to outliers}
\label{sec:outlier}

One of the important properties of the error model used in this paper
is that curves fitted to data become less sensitive to points that
depart significantly from the fitted curve (outliers) as the $r_i$
parameters of the measurements are increased.  This is a well-known
feature of models based on the Student's $t$ distribution (see, e.g.,
Ref.~\cite{bib:lange1989}).

The reduced sensitivity to outliers is illustrated in
Fig.~\ref{fig:outlier} for the case of averaging five measurements of
the same quantity (i.e., the fit of a horizontal line).  All measured
values are assigned $\sigma_{y_i}$ and $s_i$ equal to 1.0, and in
Figs.~\ref{fig:outlier}(a) and (c) they are all fairly close to the
central value of 10.  In Figs.~\ref{fig:outlier} (b) and (d) the
middle point is at 20.  In the top two plots, the $r_i$ parameters for
all measurements are taken to be $r_i = 0.01$, which is very close to
what would be obtained with an ordinary least-squares fit.  In (a) the
average is 10; in (b) the outlier causes the fitted mean to move to
12.00.  In both cases the half-width of the confidence interval is
0.63.

\setlength{\unitlength}{1.0 cm}
\renewcommand{\baselinestretch}{0.9}
\begin{figure*}[htbp]
\begin{picture}(10.0,10.3)
\put(2,5) {\includegraphics{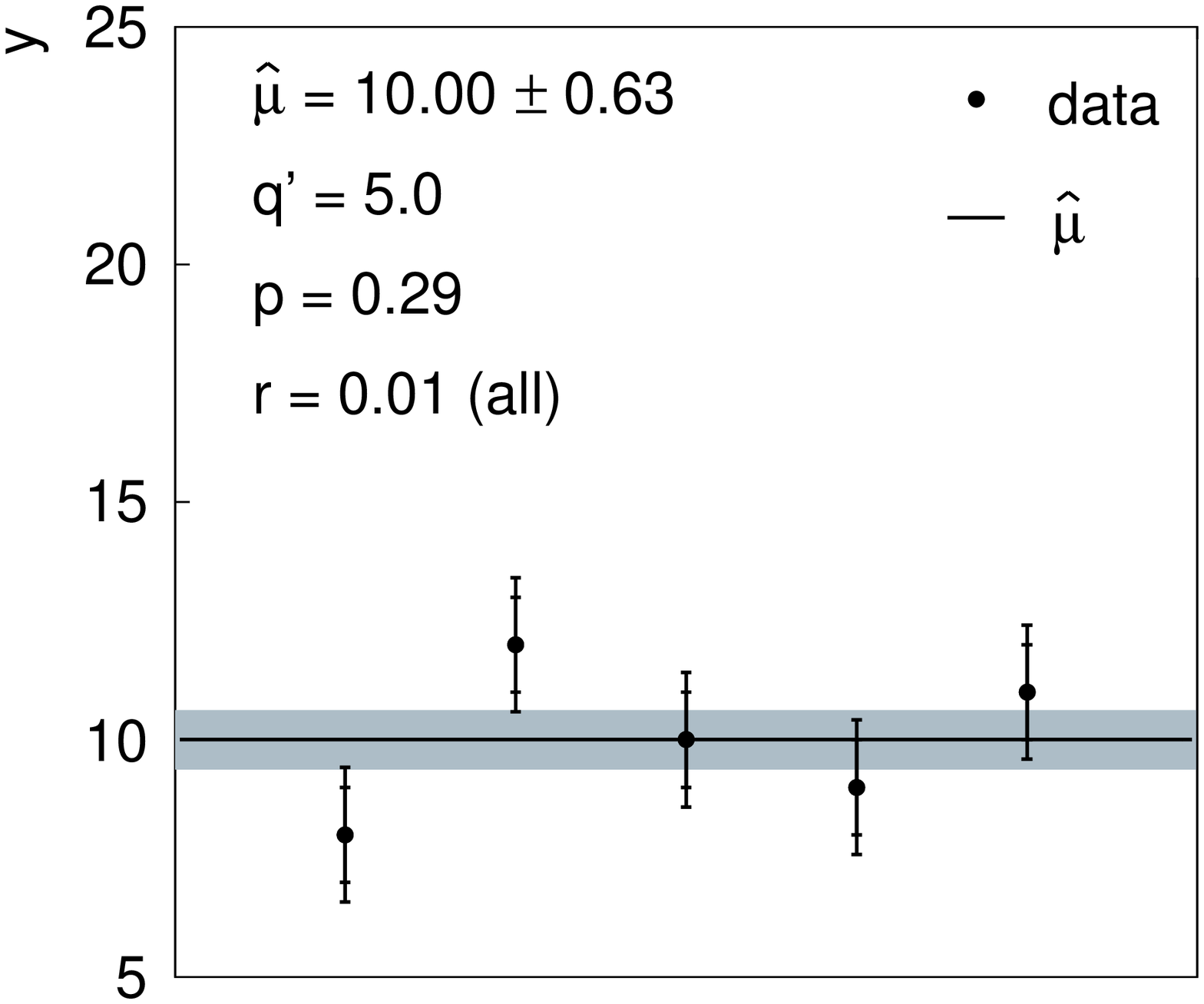}}
\put(10,5) {\includegraphics{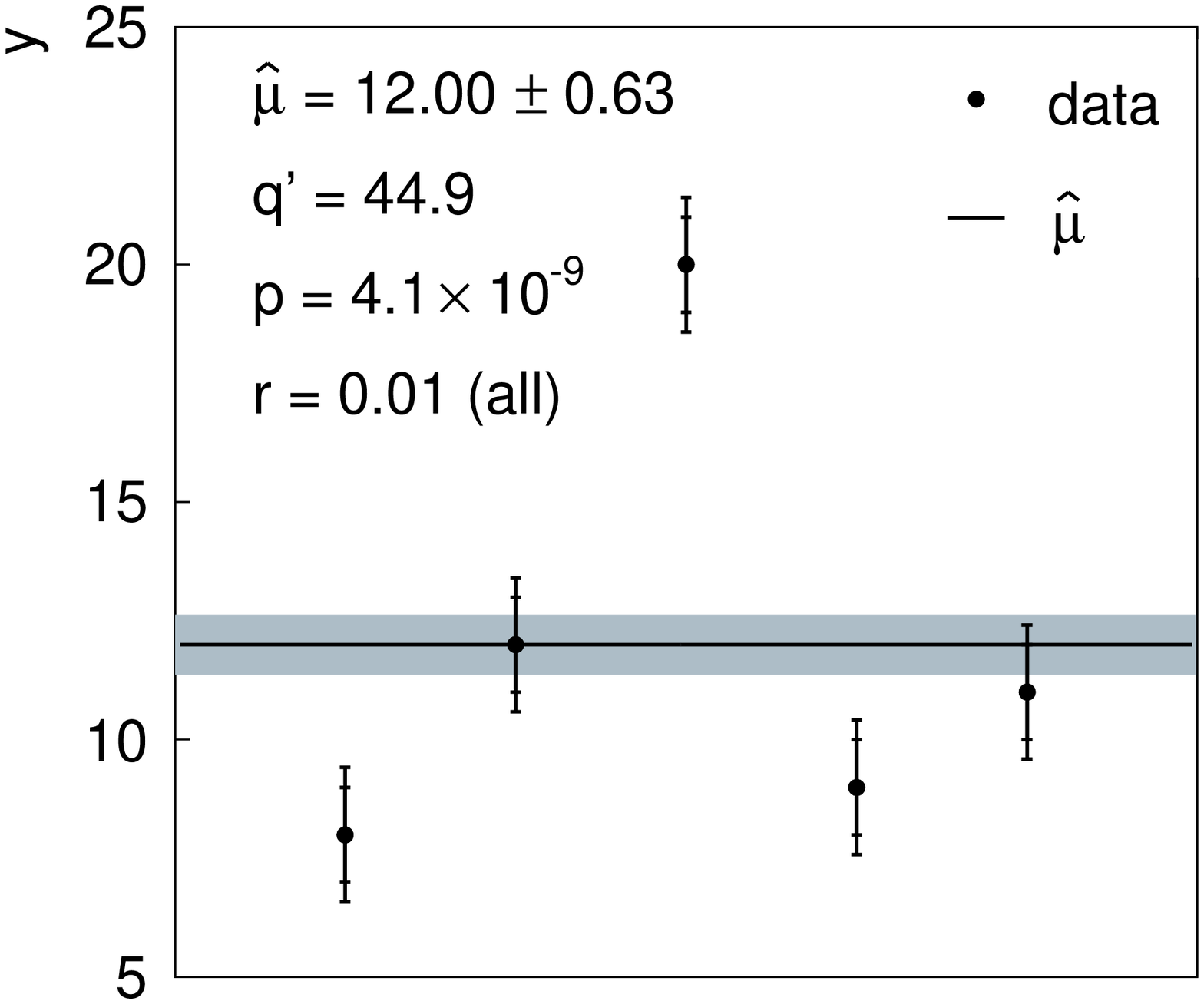}}
\put(2,0.0) {\includegraphics{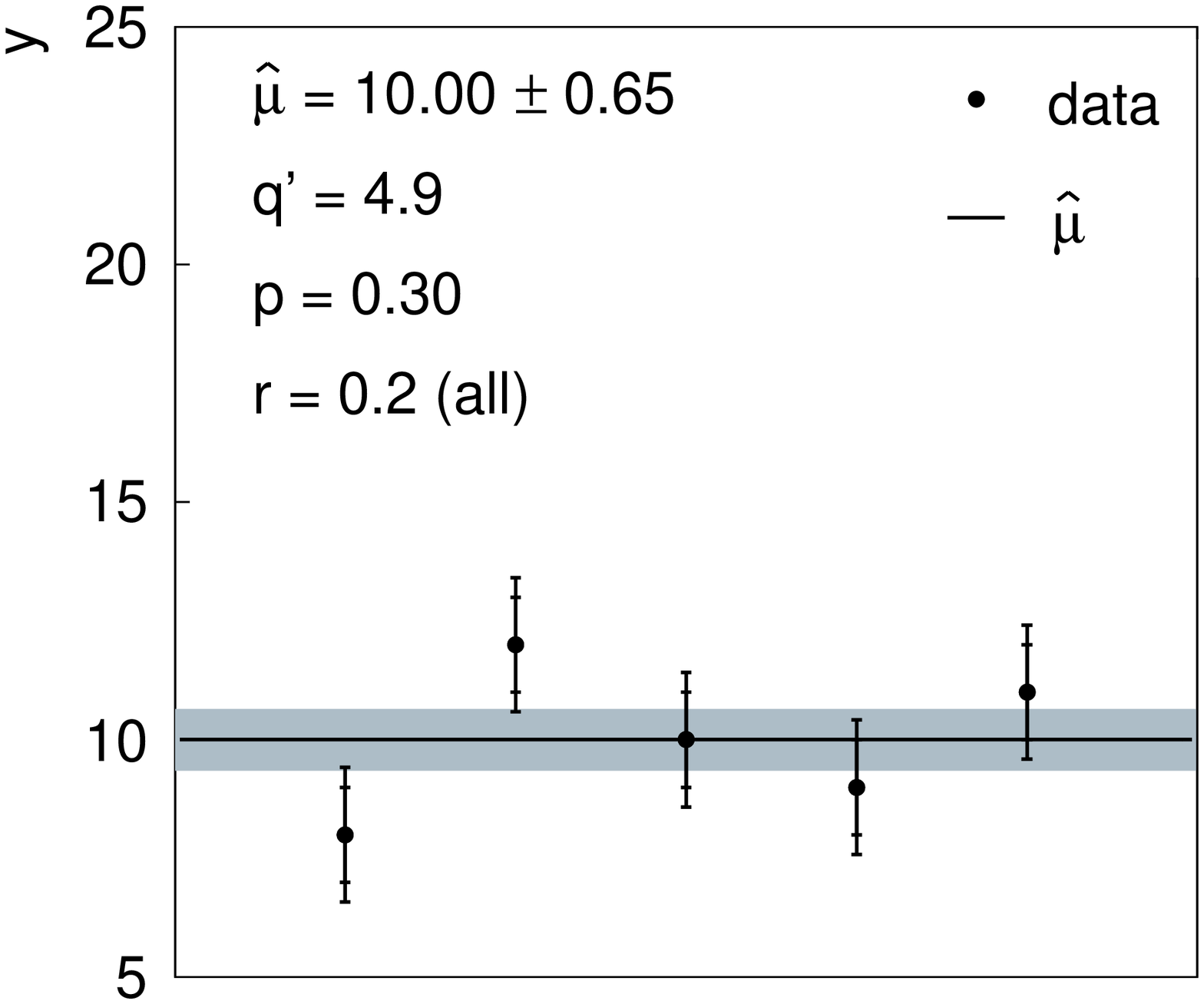}}
\put(10,0.0) {\includegraphics{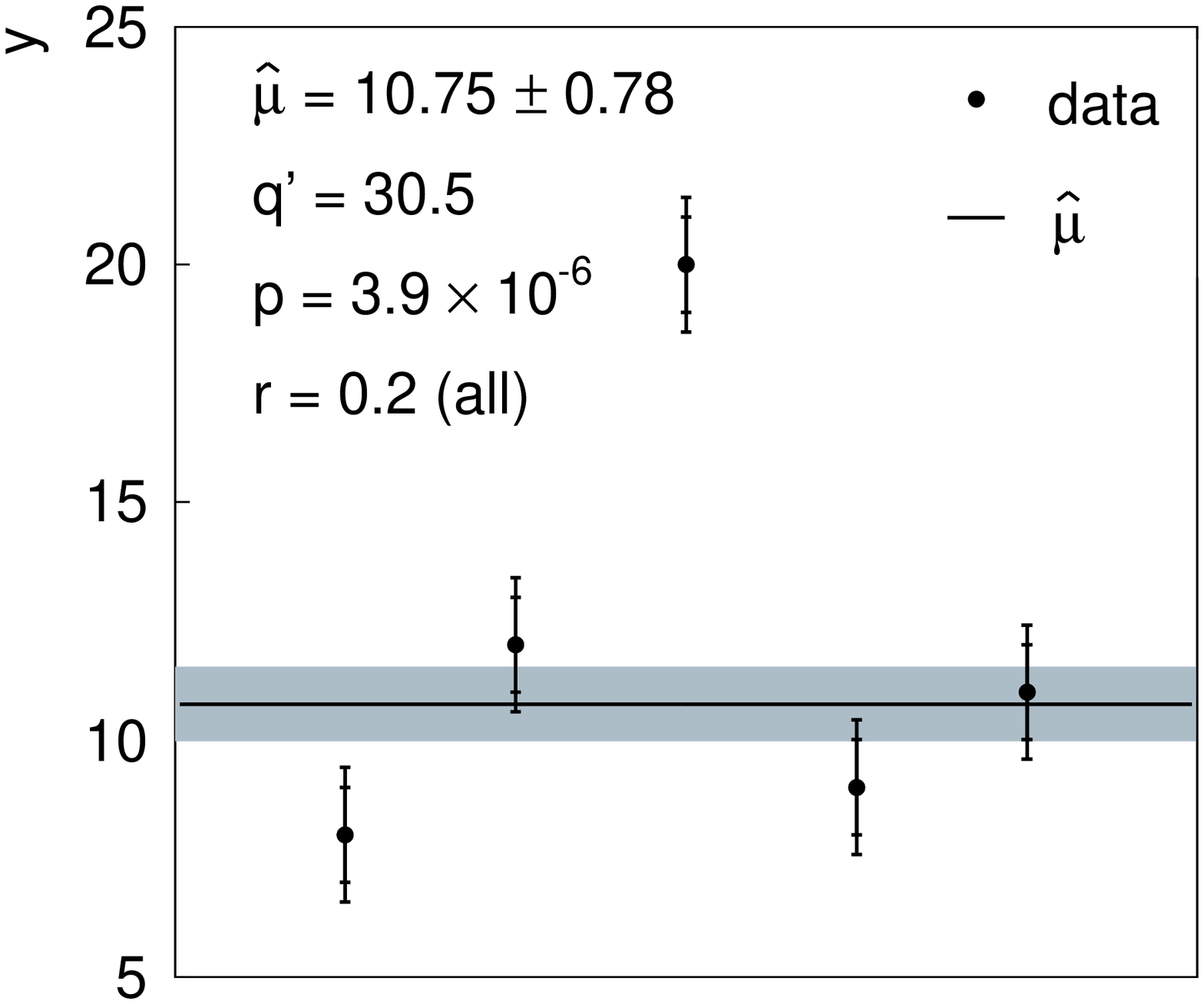}}
\put(7.8,9.5){(a)}
\put(15.8,9.5){(b)}
\put(7.8,4.5){(c)}
\put(15.8,4.5){(d)}
\end{picture}
\caption{\small Result of averaging 5 quantities: (a) no outlier, $r_i
  = 0.01$; (b) with outlier, $r_i = 0.01$; (c) no outlier, $r_i = 0.2$;
  (d) with outlier, $r_i = 0.2$.  Also indicated on the plots are the
  values of the Bartlett-corrected goodness-of-fit statistic
  $q^{\prime}$ and the corresponding $p$-value.}
\label{fig:outlier}
\end{figure*}
\renewcommand{\baselinestretch}{1}
\small\normalsize

In the lower two plots, (c) and (d), all of the points are assigned
$r_i = 0.2$, i.e., a 20\% relative uncertainty on the systematic
error.  In the case with no outlier, (c), the estimated mean stays at
10.00, and the half-width of the confidence interval only increases a
small amount to 0.65.  With the outlier in (d), the fitted mean is
10.75 with an interval half-width of 0.78.  That is, the amount by
which the outlier pulls the estimated mean away from the value
preferred by the other points (10.00) is substantially less than with
$r_i = 0.01$, (fitted mean 12.00).  Furthermore, the lower
compatibility between the measurements results in a confidence
interval that is larger than without the outlier (half-width 0.78
rather than 0.65).  When the $r_i$ are small, however, the interval
size is independent of the goodness of fit.  Both the increase in the
size of the confidence interval and the decrease in sensitivity to the
outlier represent important improvements in the inference.  It is
important to note that the above-mentioned properties pertain to the
case where each measurement has its own bias parameter $\theta_i$ with
its own $r_i$.

It might appear that one would obtain a result roughly equivalent to
that of the proposed model by using the ordinary least-squares
approach, i.e., the log-likelihood of Eq.~(53), and simply making the
replacement $\sigma_{u_i} \rightarrow \sigma_{u_i}(1 + r_i)$.  In the
example shown above with all $r_i = 0.2$, however, the result is
$\hat{\mu} = 10.00 \pm 0.70$ without the outlier (middle data point at
10) and $\hat{\mu} = 12.00 \pm 0.70$ if the middle point is moved to
20.  So by inflating the systematic errors but still using least
squares, one increases the size of the confidence interval by an
amount that does not depend on the goodness of fit and the sensitivity
to outliers is not improved.

\section{Treatment of  correlated uncertainties}
\label{sec:corr}

The phrase ``correlated systematic uncertainties'' is often taken to
mean the situation where a nuisance parameter affects multiple
measurements in a coherent way.  Suppose, for example, that the
expectation values $E[y_i]$ of measured quantities $y_i$ with $i = 1,
\ldots, L$ are functions $\varphi_i(\bvec{\mu}, \bvec{\theta})$ of
parameters of interest $\bvec{\mu} = (\mu_1, \ldots, \mu_M)$ and
nuisance parameters $\bvec{\theta} = (\theta_1, \ldots, \theta_N)$.
Suppose further that the nuisance parameters are defined such that for
$\bvec{\theta} = 0$ the $y_i$ are unbiased measurements of the nominal
model $\varphi_i(\bvec{\mu})$.  Expanding $\varphi_i$ to first order
in $\bvec{\theta}$ therefore gives

\begin{equation}
\label{eq:yidist}
E[y_i] = \varphi_i(\bvec{\mu}, \bvec{\theta}) \approx
\varphi_i(\bvec{\mu}) + \sum_{j=1}^N R_{ij} \theta_j \;,
\end{equation}

%

\noindent where the factors $R_{ij} = \left.
  \partial \varphi_i/\partial \theta_j \right|_{\bvec{\theta} = 0}$
determine how much $\theta_j$ biases the measurement $y_i$.
%

Suppose that the $R_{ij}$ are known, either from symmetry (e.g., a
particular $\theta_j$ could be known to contribute equally to all of
the $y_i$) or they are determined using a Monte Carlo simulation.  As
before suppose one has a set of independent Gaussian-distributed
control measurements $u_j$ used to constrain the nuisance parameters,
with mean values $\theta_j$ and standard deviations $\sigma_{u_j}$.
One can define the total bias of measurement $y_i$ as

\begin{equation}
\label{eq:betai2}
b_i = \sum_{j=1}^N R_{ij} \theta_j \;.
\end{equation}

\noindent and an estimator for $b_i$ is

\begin{equation}
\hat{b}_i = \sum_{j=1}^N R_{ij} u_j \;.
\end{equation}

\noindent These estimators of the biases are correlated.  As the
control measurements are assumed independent, and therefore
$\mbox{cov}[u_k, u_l] = V[u_k] \delta_{kl}$, the covariance of the
bias estimators is

\begin{equation}
  U_{ij} = \mbox{cov}[\hat{b}_i, \hat{b}_j] 
  = \sum_{k=1}^N R_{ik} R_{jk} V[u_k] \;.
\end{equation}

\noindent It is in the sense described here that the proposed model is
capable of treating correlated systematic uncertainties.  That is,
although the control measurements $u_i$ are independent they result in
a nondiagonal covariance for the estimated biases of the measurements.

The matrix $U_{ij}$ is shown here only to illustrate how correlated
bias estimates can be related to independent control measurements and
it is not explicitly needed in the type of the analysis described
here.  The full likelihood can be constructed from the measurements
$y_i$ together with their expectation values given by Eq.~(65), where
the $R_{ij}$ are assumed known.  That is, in the log-likelihood of
Eqs.~(53) or (55) the terms $y_i - \varphi(x_i; \bvec{\mu}) -
\theta_i$ are replaced by $y_i - \varphi_i(\bvec{\mu}) - \sum_{j=1}^N
R_{ij} \theta_j$.  If the variances $\sigma^2_{u_i}$ of the control
measurements $u_i$ are themselves uncertain then they are treated as
adjustable parameters with independent gamma-distributed estimates.



\section{Discussion and conclusions}
\label{sec:conc}

The statistical model proposed here can be applied in a wide variety
of analyses where the standard deviations of Gaussian measurements are
deemed to have a given relative uncertainty, reflected by the
parameters $r_i$ defined in Eq.~(\ref{eq:rsigmav}).  The quadratic
constraint terms connecting control measurements to their
corresponding nuisance parameters that appear in the log-likelihood
are replaced by logarithmic terms (cf.\ Eqs.~(\ref{eq:lnpxu}) and
(\ref{eq:proflikmub})).  The resulting model is equivalent to taking a
Student's $t$ distribution for the control measurements, with the
number of degrees of freedom given by $\nu = 1/2r^2$.

It is not uncommon for systematic errors, especially those related to
theoretical uncertainties, to be uncertain themselves to several tens
of percent.  The model presented here allows such uncertainties to be
taken into account and it has been shown that this has interesting and
useful consequences for the resulting inference.  
Confidence intervals are found to increase in size if the goodness of
fit is poor and can decrease slightly if the data are more internally
consistent than expected, given the level of statistical fluctuation
assumed in the model.  Averages and fitted curves become less
sensitive to outliers.

If the relative uncertainty on the systematic errors is large
enough ($r$ greater than around 0.2 in the examples studied), then the
sampling distribution of likelihood-ratio test statistics starts to
depart from the asymptotic chi-squared form.  Thus one cannot in
general apply asymptotic results for $p$-values and confidence
intervals without taking some care to ensure their validity.  In some
cases Bartlett-corrected statistics can be used; alternatively one may
need to determine the relevant distributions by Monte Carlo
simulation.  

In reporting results that use the procedure presented here it is
important to communicate all of the $r_i$ parameters.  To allow for
combinations with other measurements one should ideally report the
full likelihood, including the $r_i$ values, to permit a consistent
treatment of uncertainties common to several of the measurements.

The point of view taken here has been that the analyst must determine
reasonable values for the relative uncertainties in the systematic
errors.  One should not, for example, decide to use the proposed model
only if the goodness of fit is found to be poor. Rather, the $r_i$
parameters should reflect the accuracy with which the systematic
variances have been estimated and the resulting inference about the
parameters of interest then incorporates this knowledge in a manner
that is valid for any data outcome.

An alternative mentioned here as a possibility would be to fit a
common relative uncertainty to all systematic errors (a global $r$),
e.g., when averaging a set of numbers for which no $r$ values have
been reported.  This is analogous to the scale-factor procedure used
by the Particle Data Group \cite{bib:PDG} or the method of DerSimonian
and Laird \cite{bib:dersim} widely used in meta-analysis.  Note,
however, that in arriving at the log-likelihood (\ref{eq:proflikmub}),
a number of terms dependent on the $r_i$ were dropped, as they were
considered fixed constants.  If the $r_i$ are adjustable parameters
then these terms, given in App.~\ref{sec:proflnL}, must be retained in
the log-likelihood.

\section*{Acknowledgements}


Many thanks for stimulating discussions and useful assistance are due
to Lorenzo Moneta, Bogdan Malaescu, Nicolas Berger, Francesco Span\`o,
Adam Bozson, Nicolas Morange and numerous members of the ATLAS
Collaboration.  Many suggestions related to this work were obtained at
the 2018 Workshop on Advanced Statistics for Physics Discovery at the
University of Padova, supported by the Marie-Curie ITNs
AMVA4NewPhysics and INSIGHTS, in particular from David van Dyk,
Alessandra Brazzale and Bodhisattva Sen.  This work was supported in
part by the U.K.\ Science and Technology Facilities Council.

\appendix

\section{Exact relation between the $r$ parameter and the relative
error on the error}
\label{sec:rdef}

The parameter $r$ was defined in Eq.~(\ref{eq:rsigmav}) as

\begin{equation}
\label{eq:rdef}
r = \frac{1}{2} \frac{\sigma_{v}}{E[v]} \,,
\end{equation}

\noindent where we drop the subscript $i$ as we are focusing on a
single measurement.  Here $v$, the estimate of a variance
$\sigma_u^2$, is assumed to follow a gamma distribution with
expectation value $E[v] = \sigma_u^2$.

The physicist is more likely to work with the estimated standard
deviation rather than the variance, i.e., with $s = \sqrt{v}$.  From
error propagation we have that the standard deviation of $s$ is

\begin{equation}
\sigma_s \approx \left| \frac{ ds}{dv} \right|_{v = \sigma_u^2} \sigma_v
= \frac{1}{2} \frac{\sigma_s^2}{\sigma_u}
\end{equation}

\noindent If we approximate the $E[s] \approx (E[v])^{1/2} =
\sigma_u$, then the relative uncertainty on the standard deviation is

\begin{equation}
\label{eq:relerr}
\frac{\sigma_s}{E[s]} \approx \frac{1}{2} \frac{\sigma_v}{\sigma_u^2}
= \frac{1}{2} \frac{\sigma_v}{E[v]} \,.
\end{equation}

\noindent Equation~(\ref{eq:relerr}), based on linear error
propagation, holds to the extent that the nonlinearity of $v = s^2$ is
not large over the range $v = E[v] \pm \sigma_v$.  For sufficiently
large $r$, however, this assumption will break down and one can
no longer interpret $r$ as a relative error on the error.  

Starting from a gamma distribution (\ref{eq:gammapdf}) with parameters
$\alpha$ and $\beta$ for the distribution of $v$, the pdf of $s =
\sqrt{v}$ is given by

\begin{equation}
  g(s | \alpha, \beta) = \left| \frac{dv}{ds} \right| f(v(s)|\alpha, \beta)
  = \frac{2 \beta^{\alpha}}{\Gamma(\alpha)} s^{2 \alpha - 1} e^{- \beta s^2} \,,
\end{equation}

\noindent where $\alpha = 1/4r^2$ and $\beta = \alpha / \sigma_{u}^2$.
This is a special case of the Nakagami distribution
\cite{bib:Nakagami,bib:NakagamiWiki}, which has mean and variance

\begin{eqnarray}
E[s] & = & \frac{\Gamma(\alpha + \frac{1}{2})}{\Gamma(\alpha) \sqrt{\beta}} 
\,, \\*[0.3 cm]
V[s] & = & \frac{\alpha}{\beta} - \frac{1}{\beta} \left(
\frac{\Gamma(\alpha + \half)}{\Gamma(\alpha)} \right)^{2} \,.
\end{eqnarray}

\noindent  The exact relative uncertainty in the standard
deviation is

\begin{equation}
\label{eq:rs}
r_s \equiv \frac{\sqrt{V[s]}}{E[s]}
= \sqrt{  \alpha \left( \frac{\Gamma(\alpha)}
{\Gamma(\alpha + \half)} \right)^2
- 1} \,,
\end{equation}

\noindent which is shown in Fig.~\ref{fig:rs}.
For example, $r = 1$ gives $r_{s} = 1.09$.  Thus for  relevant
values of $r$ one can safely approximate $r_s \approx r$.

\setlength{\unitlength}{1.0 cm}
\renewcommand{\baselinestretch}{0.9}
\begin{figure}[htbp]
\begin{picture}(6.0,6.0)
\put(0.5,-0.5){\includegraphics{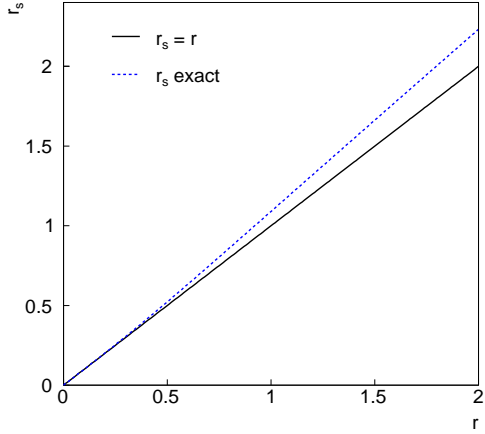}}
\end{picture}
\caption{\small  The exact relative uncertainty
$r_s$ as a function of the parameter $r$ (see text).}
\label{fig:rs}
\end{figure}
\renewcommand{\baselinestretch}{1}
\small\normalsize

\section{Derivation of the profile log-likelihood}
\label{sec:proflnL}

The full log-likelihood for the gamma error model from
Eq.~(\ref{eq:lnLfull}), written here with the constant terms, is

\vspace{-0.5 cm}

\begin{eqnarray}
\label{eq:lnLfull2}
\ln L(\bvec{\mu}, \bvec{\theta}, \bvec{\sigma}_{\bvec{u}}^2) & = & 
\ln P(\bvec{y} | \bvec{\mu}, \bvec{\theta}) 
\nonumber \\*[0.3 cm]
& - & \frac{1}{2}
\sum_{i=1}^N \left[
\frac{(u_i - \theta_i)^2}{\sigma_{u_i}^2} 
+   \ln \sigma_{u_i}^2 + \ln (2 \pi) \right] \nonumber \\*[0.3 cm]
& + & \sum_{i=1}^N \left[ \alpha_i \ln \beta_i - \ln \Gamma(\alpha_i) \right.
\nonumber \\*[0.3 cm]
& + & \left. (\alpha_i - 1) \ln v_i - \beta_i v_i \right] \,,
\end{eqnarray}

\noindent where the parameters of the gamma distribution $\alpha_i$
and $\beta_i$ are related to $r_i$ and $\sigma_{u_i}^2$ by
Eqs.~(\ref{eq:alphai}) and (\ref{eq:betai}).  By using the profiled
values for $\widehat{\widehat{\sigma^2}}_{u_i}$ from
Eq.~(\ref{eq:sigmau2hathat}) we obtain

\begin{eqnarray}
\label{eq:proflikmubfull2}
& & \ln L^{\prime}(\bvec{\mu}, \bvec{\theta}) = 
\ln L(\bvec{\mu}, \bvec{\theta}, 
\widehat{\widehat{\sigma^2}}_{\bvec{u}}(\bvec{\theta})) 
\nonumber \\*[0.3 cm]
& = & \ln P(\bvec{y} | \bvec{\mu}, \bvec{\theta}) 
 -  \frac{1}{2} \sum_{i=1}^N \left[ 
\frac{(1 + 2 r_i^2) (u_i - \theta_i)^2}{v_i + 2 r_i^2 (u_i - \theta_i)^2}
\right. \nonumber \\*[0.3 cm]
& + & \left. 
\ln \left(\frac{v_i + 2 r_i^2 (u_i - \theta_i)^2}{1 + 2 r_i^2} \right)
+ \ln (2 \pi)
\right] \nonumber \\*[0.3 cm]
& + & \sum_{i=1}^N \left[ \frac{1}{4 r_i^2} 
\ln \frac{1 + 2 r_i^2}{4 r_i^2 \left[ v_i + 2 r_i^2 (u_i - \theta_i)^2 \right] }
- \ln \Gamma \left( \frac{1}{4 r_i^2} \right) \right. \nonumber \\*[0.3 cm]
& + & \left. \left( \frac{1}{4 r_i^2} - 1 \right) \ln v_i
- \frac{ v_i (1 + 2 r_i^2) }
{ 4r_i^2 \left[ v_i + 2 r_i^2 (u_i - \theta_i)^2 \right] }
 \right] \,.
\end{eqnarray}

\noindent By rearranging terms the profile likelihood can be written (cf.\
Eq.~(\ref{eq:proflikmub}))

\begin{eqnarray}
\label{eq:proflikmub2}
\ln L^{\prime}(\bvec{\mu}, \bvec{\theta}) 
& = & \ln P(\bvec{y} | \bvec{\mu}, \bvec{\theta})
 \\*[0.3 cm]
& - &  \frac{1}{2} \sum_{i=1}^N 
\left( 1  + \frac{1}{2 r_i^2} \right) \ln \left[
1 + 2 r_i^2 \frac{(u_i - \theta_i)^2}{v_i} 
\right] + C \,, \nonumber
\end{eqnarray}

\noindent where

\begin{eqnarray}
\label{eq:cofr}
C & = & - \frac{1}{2} \sum_{i=1}^N \left[
\left(1 + \frac{1}{2 r_i^2} \right) \left( 1 + \ln
\frac{v_i}{1 + 2 r_i^2} \right) +
\frac{1}{2 r_i^2} \ln ( 4 r_i^2 ) \right. \nonumber \\*[0.3 cm]
& + & \left.
2 \ln \Gamma \left( \frac{1}{4 r_i^2} \right) 
+ \left(2 - \frac{1}{2 r_i^2}  \right) \ln v_i  + \ln (2 \pi) \right]
\end{eqnarray}

\noindent does not depend on any of the adjustable parameters of the
problem and thus can be dropped.  If, however, one were to treat the
$r_i$ as free parameters then $C$, or at least those terms depending
on the $r_i$, must be retained.

\end{document}